# Atomic-Scale Observation of Moiré potential in Twisted Hexagonal Boron Nitride Layers by Electron Microscopy

Rina Mishima,[†] Takuro Nagai,[§] Hiroyo Segawa, Masahiro Ehara,[‡] and Takashi Uchino[†*]

[†]Department of Chemistry, Graduate School of Science, Kobe University, Nada, Kobe 657-8501, Japan

[§]Research Network and Facility Services Division, National Institute for Materials Science, Tsukuba, Ibaraki 305-0044, Japan

[ǁ]Research Center for Electronic and Optical Materials, National Institute for Materials Science, Tsukuba, Ibaraki 305-0044, Japan

[‡]Research Center for Computational Science, Institute for Molecular Science, Okazaki, Aichi, 444-8585, Japan

**ABSTRACT:** Moiré superlattices (MSLs) are an emerging class of two-dimensional functional materials whose electronic states can be tuned by the twist angle between two van der Waals layers and/or the relative placement of the layers. The intriguing properties of MSLs are closely correlated to the moiré potential, which is the electrostatic potential induced by interlayer coupling. Intensive efforts have been made to understand the nature and distribution of the moiré potential by using various experimental and theoretical techniques. However, the experimental observation of the moiré potential is still challenging because of the possible presence of the surface and/or interlayer contaminants. In this work, we develop a method to obtain hexagonal boron nitride (hBN) nanolayers (with or without twist) using a specially designed chemical exfoliation technique. The resulting hBN nanolayers are atomically clean and strain free, hence providing ideal MSLs for the investigation of their moiré potential. Aberration-corrected high resolution transmission electron microscopy measurements on the twisted hBN nanolayers allow us to observe moiré diffraction spots in Fourier space. Then, the moiré potential is reconstructed by the inverse fast Fourier transform of the moiré diffraction spots. It has been revealed that the local interlayer atomic overlap plays a decisive role in determining the periodicity and distribution of the moiré potential, as supported by density functional theory calculations. This work not only provides a general strategy to observe the moiré potential in MSLs, but it also expands the application of electron microscopy to the further study of MSLs with atomic resolution.

## ■ INTRODUCTION

Vertically stacked twisted layers of two-dimensional (2D) materials provide a fruitful ground for the exploration of unique and novel quantum phenomena in van der Waals (vdW) material systems.[1–14] Stacking of 2D vdW materials with a small twist angle $\theta$ ($\theta$<~10°) and/or a slight lattice mismatch results in the reconstructed moiré superlattices (MSLs), accompanied by the enhanced interlayer coupling.[4,14–16] These reconstructed MSLs especially with magic twist angles around $\theta\approx 1.1°$ can host flat electronic bands to form highly correlated phases, even showing superconductivity.[1,8,9] In the small angle regime, the commensuration cell lattice constant can be equal to the moiré periodicity $D$, which is given by[15]

$$D = \frac{a}{2\sin\frac{\theta}{2}} \quad (1),$$

where $a$ is the lattice constant of the constituent layer. For large twist angles ($\theta$>~15°), on the other hand, the commensuration cell lattice constant is generally much greater than $D$.[15] However, there exist a few special twist angles at, e.g., 13.2° and 21.8° for twisted hexagonal bilayers,[15,17,18] where the "commensuration periodicity" is equal to $D$. Thus, from the view point of MSLs, any large twist angles except for the above special twist angles can be regarded as incommensurate twist angles. In the large angle regime, it is believed that atomic reconstruction hardly occurs, meaning that a rigid lattice model is a reasonable representation of the actual system.[4] Although the interlayer coupling is supposed to be suppressed in these incommensurate MSLs,[19] it has recently been revealed that this is not necessarily the case for 30°-twisted vdW bilayers, forming quasicrystalline patterns with a 12-fold rotational symmetry.[20–22] In these quasicrystalline layers, two twisted layers are coupled via Umklapp electron-electron scattering, which results in rich electronic structures, such as mirrored Dirac cones in graphene quasicrystals[20] and mini-gaps near the valence band maximum in tungsten diselenide quasicrystals.[23] Thus, incommensurate MSLs and related moiré potentials have attracted increasing attention,[24,25] offering an interesting experimental platform to explore moiré physics beyond the small twist-angle regime.[26,27] However, atomic-scale visualization of moiré potential is highly technically demanding. At present, scanning tunneling microscopy/spectroscopy (STM/STS) is one possible technique available for that purpose.[23,28,29]

Here we use an aberration-corrected high-resolution transmission electron microscopy (HRTEM) technique as a tool to explore the moiré potential in atomic scale. Although HRTEM has been often employed to obtain atomic resolution imaging of 2D materials and related MSLs,[30–35] we show that this technique is also applicable for imaging the moiré potentials, which are reconstructed from the moiré diffraction spots in Fourier space. In previous works, the diffraction spots due to moiré potential have not been explicitly observed by HRTEM because of the general presence of hydrocarbon contaminants which are often trapped during the assembly of 2D stacked materials.[36–38] These contaminants yield a severe background noise in Fourier space, obscuring weak moiré diffraction spots. Furthermore, any interlayer contaminants will modify the interlayer distance, leading to a fluctuation of moiré potential.[39,40] In this work, we overcome the problem by developing a specially designed chemical exfoliation method to obtain atomically clean MSLs, which are ideal for the investigation of the moiré structure and the related potential by HRTEM technique.

## ■ EXPERIMENTAL SECTION

We develop a method to exfoliate atomically clean and strain-free hBN nanolayers from hBN particles. It has been well recognized that hBN nanostructures are characterized by their excellent thermal and chemical stability and unique electronic and optical properties,[41] providing an ideal building block for the investigation of vdW potentials.[42] Our exfoliation procedure is based on an intercalation-based method;[43] however, unlike conventional methods, it does not require any sonication, sharing nor electrochemical process as a driving force, hence yielding mechanically intact and strain-free samples. Briefly, we start by creating hBN/$H_2SO_4$ intercalation compounds by heating hBN powders in an aqueous $H_2SO_4$ solution.[44,45] The resulting solution is neutralized by a solution of $NaHCO_3$ to form sodium sulphate salt in between the hBN layers. We found that this acid/base reaction leads to spontaneous exfoliation of high-quality hBN layers suitable for the characterization of moiré potentials by HRTEM (see the Supporting Information and Fig. S1 for details).

HRTEM and scanning TEM (STEM) observations were carried out on a JEOL JEM-ARM300F instrument equipped with a cold field emission electron gun and a spherical aberration (Cs) corrector at an acceleration voltage of 80 kV, under $1 \times 10^{-5}$ Pa in the specimen column. We analyzed strain of the exfoliated hBN layers based on the HRTEM images using the Peak Pairs Analysis (PPA) software package for Gatan Digital Micrograph.[46]

In order to calculate the interlayer electrostatic potential, we performed quantum chemical calculations using density functional theory (DFT) methods implemented in the Gaussian 16 suite of programs.[47] In this work, hydrogen terminated clusters with various sizes and two different edges (zigzag and armchair) were used to model the local structure of twisted hBN bilayers with different twist angles.

The morphology of the exfoliated layers was also observed using a dynamic force mode atomic force microscope (AFM; SII SPA400).

Further details about the characterization and calculational procedures used in the work can be found in the Supporting Information.

## ■ RESULTS AND DISCUSSION

**General characteristics of exfoliated hBN.** Figure 1a shows a typical AFM image of the exfoliated hBN sheets with a lateral size of a few μm. Each sheet has a relatively smooth surface with a thickness *t* of 2–4 nm. Considering that the interlayer distance of hBN is ~0.33 nm, we can estimate that each nanosheet consists of ~5–15 BN layers. We show in Fig.2a a low-magnification annular dark field scanning TEM (ADF-STEM) image of the exfoliated hBN, showing a random stacking of micrometer-sized hBN layers. A high-magnification ADF-STEM image of non-twisted hBN layers is given in Fig. 2b, illustrating 2D honeycomb structures of BN with asymmetric ADF scattering intensities. One may expect that three boron and three nitrogen atoms in each hexagonal ring are responsible for this asymmetric contrast, which can be in principle distinguished by their intensity based on the Z-(atomic number) contrast principle.[48] However, we should note that our exfoliated hBN nanosheets are likely to have more than 5 AA′-stacked BN layers, which are formed by stacking many anti-aligned layers B to N and N to B. In such layers, the expected Z-contrast will be averaged out to yield symmetric ADF scattering intensities.[49] We hence assume that observed asymmetric ADF contrast results from a slight specimen tilt. To confirm the assumption, we performed simulations by changing *x*-tilt angle φ (for details, see Fig. S2) using a comprehensive multi-purpose crystallographic program *Recipro*.[50] After several try-and-error attempts, we found that the specimen of $t$ = ~8 nm and φ = ~1° yields the asymmetric contrast which is most comparable to that of the observed ADF-STEM image (see Fig. 2c). Although the thickness estimated from the simulation is somewhat greater than that obtained from an AFM image shown in Fig. 1, these estimated values are reasonably consistent with each other. Thus, we can conclude that the thickness of our exfoliated hBN sheets lies in the range from ~3 to ~8 nm, which will be thin enough to satisfy the weak phase object approximation (WPOA) especially for the samples consisting of light elements such as B and N. Under WPOA, the TEM image operated in the phase contrast mode at optimum focus directly reveals the projected potential.[51]

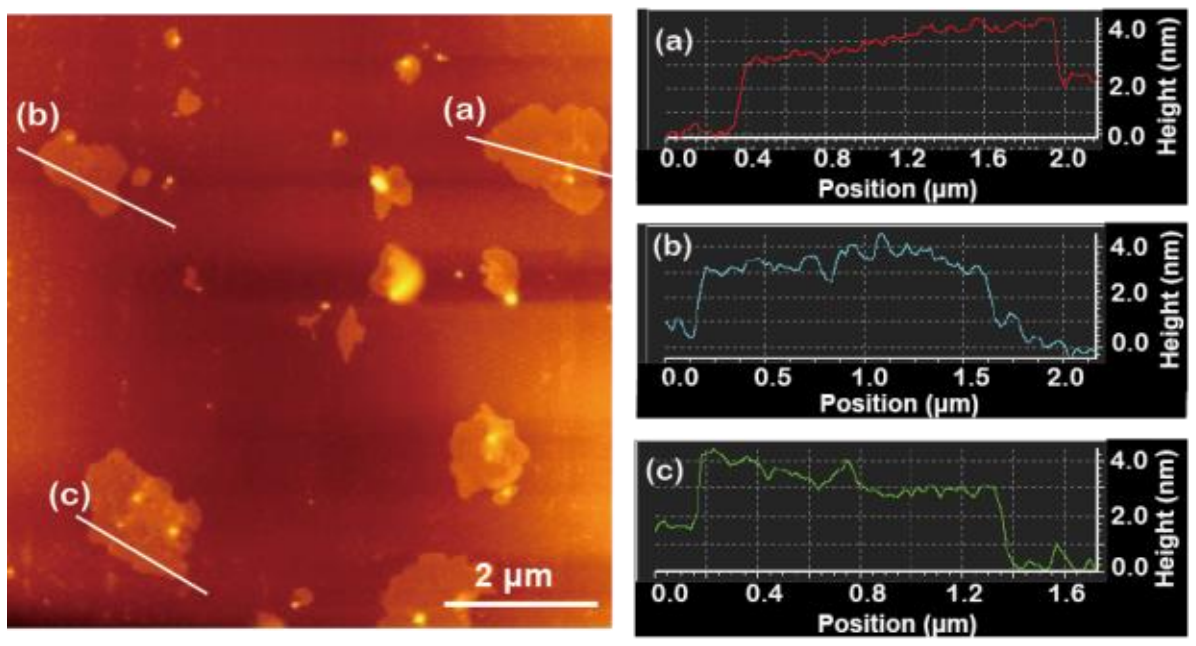

**Figure 1.** A typical AFM image of the exfoliated hBN sheets and the corresponding thickness profiles.

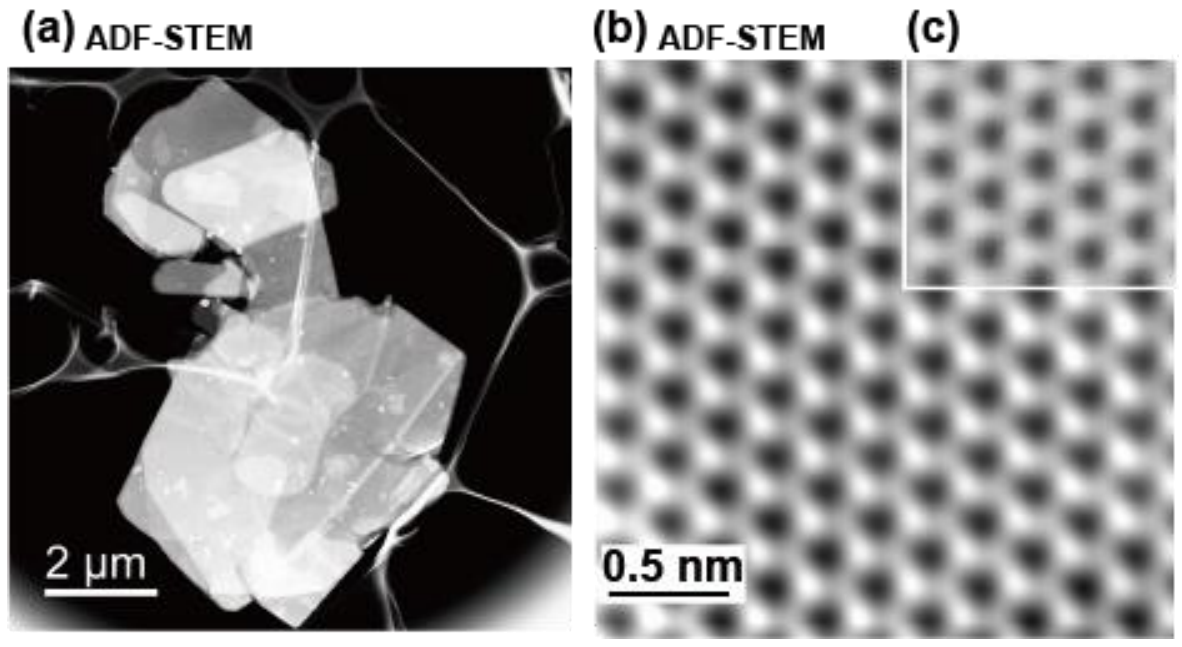

**Figure 2.** (a) Low- and (b) high-magnification ADF-STEM images of exfoliated hBN layers. (c) A simulated ADF-STEM image of a 8 nm thick hBN, which is *x*-tilted by 1° (17.5 mrad).

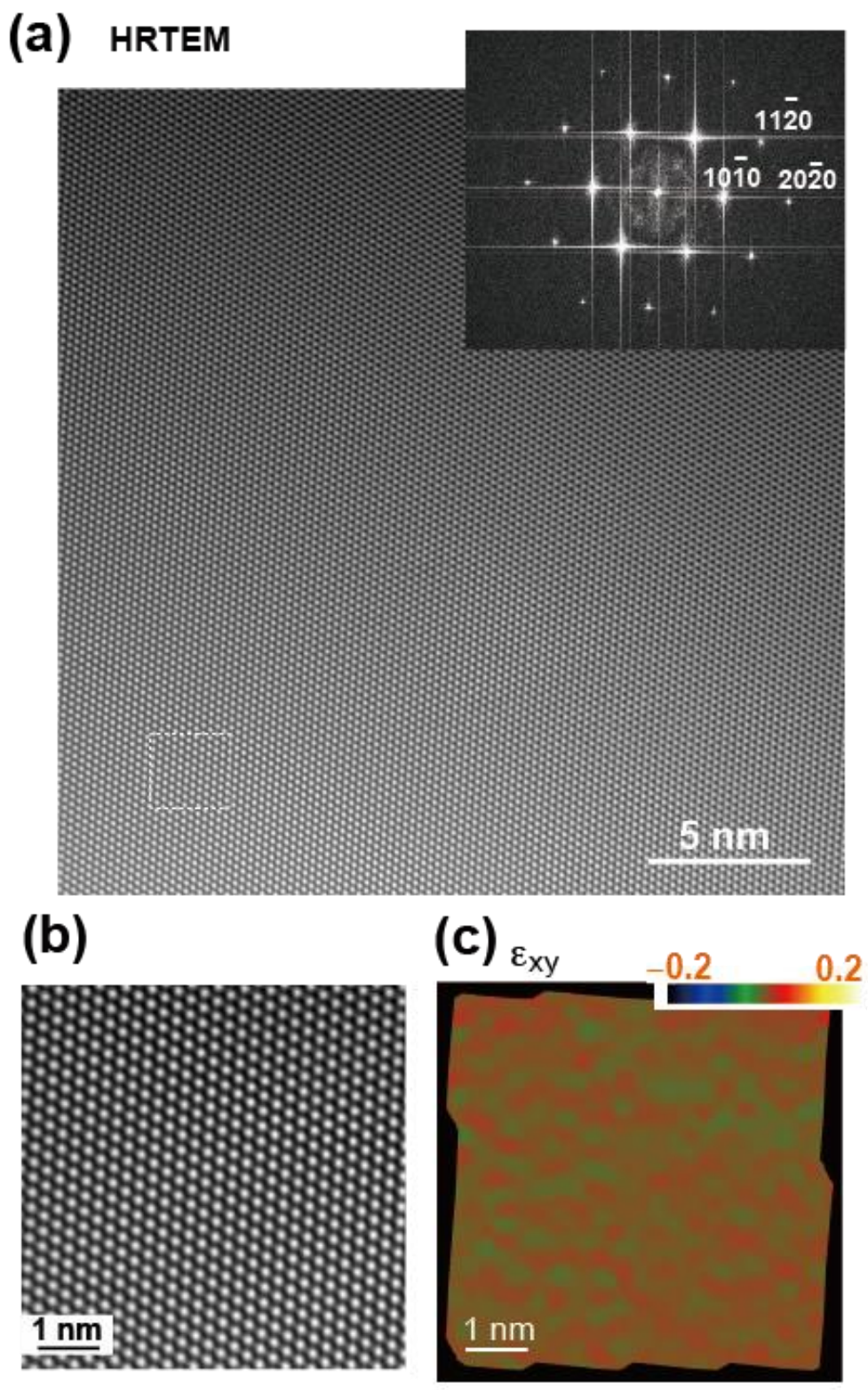

**Figure 3.** (a) An HRTEM image showing a dark contrast for the B and N atoms. The corresponding FFT image is given in inset. (b) A magnified HRTEM image of the area marked by the white, dashed square in (a). Shear strain component $\varepsilon_{xy}$ of the lattice strain tensor obtained from the HRTEM image shown in (b).

HRTEM images shown in Fig. 3a,b demonstrate that the resulting layers have a clean and adsorbates-free area extending several tens of nanometers (see also Fig. S3). The corresponding fast Fourier transform (FFT) image yields six spots of 0.21, 0.12 and 0.11 nm spacings attributed, respectively, to $10\bar{1}0$, $11\bar{2}0$ and $20\bar{2}0$ Bragg peaks (see the inset in Fig. 3a). Figure 3c is the PPA strain mapping obtained for the HRTEM image shown in Fig. 3b. As shown in Fig. 3c, the strain tensor is estimated to be as low as 1 % over the entire region of interest (see Fig. S4 for more details of the PPA analysis).

Figure 4a shows the edge area with four BN layers (see also the line profile in Fig. 4b). Any moiré patterns and the related rotational stacking faults are not present at the edge, indicating that the edges are not folded back bur rather exhibit the smallest possible scroll, as often observed in the edge structures of high-quality free-standing graphene layers.[52]

All the above results demonstrate that as far as the non-twisted hBN layers are concerned, atomically clean, flat and strain-free samples can be obtained by the present exfoliation procedure.

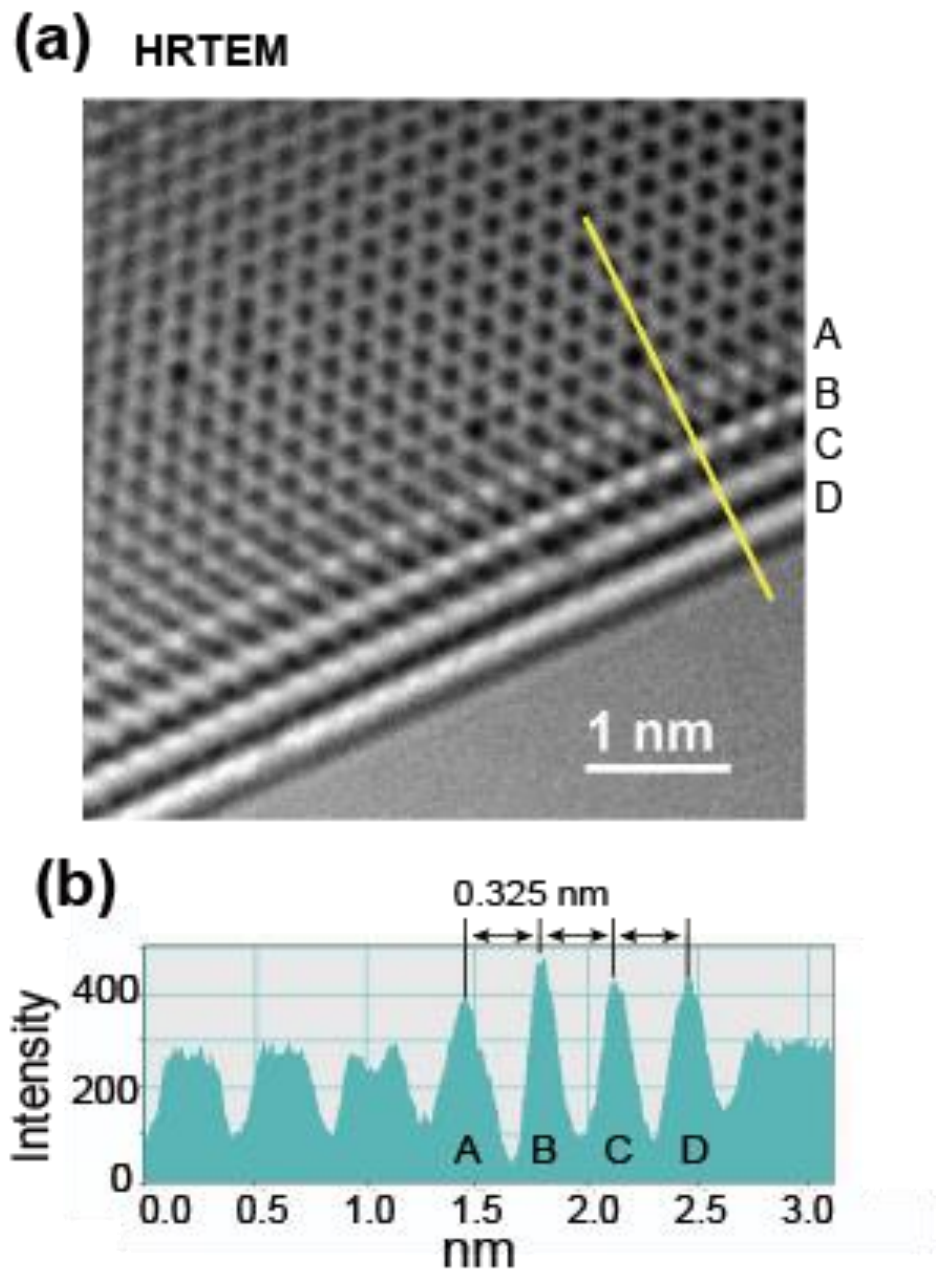

**Figure 4.** (a) An HRTEM image of the edge of an hBN multilayer. (b) Line profile as a function of distance along the yellow line in (a), showing a four-layer structure with a fringe separation of 0.325 nm.

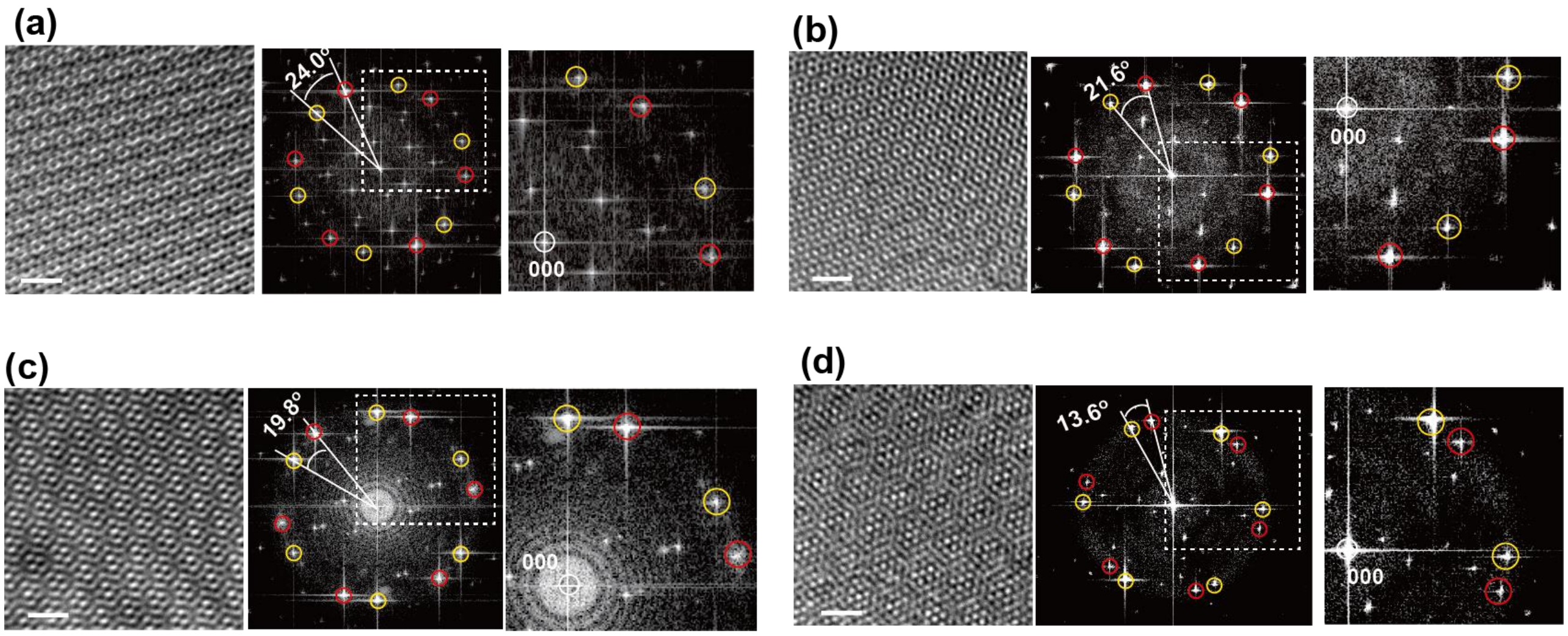

**Figure 5.** HRTEM observations on twisted hBN layers with different twist angles. (a-d) (left panel) HRTEM image. Scale bar, 1 nm. (middle panel) FFT image, showing the twist angle determined from the angle between the two closer spots. Red and yellow circles show two sets of six 100 diffraction spots belonging respectively to different layers. (right panel) A magnified FFT image of the area marked by the white dashed square shown in the middle panel.

**TEM observations on twisted hBN layers.** In addition to the non-twisted regions, we also found several contaminant-free twisted regions at different twist angles. We show in Fig. 5a-d the HRTEM and the corresponding FFT images of the twisted hBN layers with four different twist angles of $\theta$= 24.0º, 21.6º, 19.8º and 13.6 °. In this work, the twist angle $\theta$ was defined as the angle between the two closest $10\bar{1}0$ spots, as often done in previous studies. The uncertainty of $\theta$ is within the range of ±0.4º. Note, however, that the alternative twist angle $\bar{\theta}$ can be defined as the angle between the two next-closest $10\bar{1}0$ spots, i.e., $\bar{\theta} = 60° - \theta$. Since we cannot identify which definition is appropriate from the diffraction pattern alone, the selection of $\theta$ is just for convenience. We will show that this selection is not crucial for the discussion of this work in a later subsection..

The observed HRTEM images illustrate a variety of moiré patterns typical to twisted bilayers with large twist angles. The two of the observed twist angles (13.6 ° and 21.6°) are comparable to two of the commensurate angles of twisted hexagonal bilayers (13.2° and 21.8°)[17,18] within the experimental uncertainty. Thus, in what follows, we will refer to the samples with $\theta$=13.6 ° and 21.6° as nearly commensurate layers, whereas the rest of the samples as incommensurate ones. Note, however, that even in the incommensurate cases, the partially eclipsed interlayer atomic overlap, i.e., N nearly on N (or B nearly on B) configurations, is expected to be realized in the rigid lattice picture, as illustrated in Fig. S5.

The FFT images given in Fig. 5 are especially worth noting. We see that in addition to the two sets of $10\bar{1}0$ Bragg spots defining a twist angle, there exist several extra spots especially in the inner region of the $10\bar{1}0$ spots, which are located symmetrically with respect to the center spot. These extra spots have not been explicitly reported in previous HRTEM observations on MSLs.[30–35] We also found that these extra spots are not observed in the TEM image of the twisted hBN layers with hydrocarbon adsorbates (Fig. 6). This allows us to expect that clean and contaminant-free samples are necessary for the observation of the extra diffraction spots.

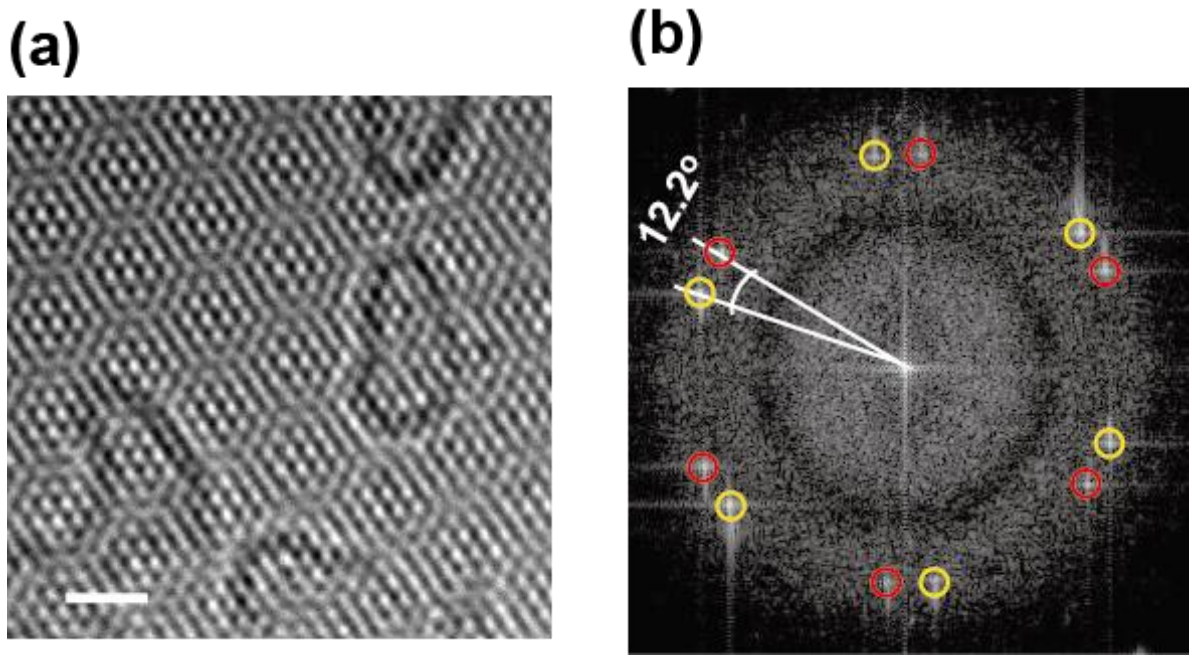

**Figure 6.** (a) HRTEM image of twisted hBN layers with hydrocarbon adsorbates. Scale bar, 1 nm. (b) Corresponding FTT image, along with the twist angle determined from the angle between the two closest spots. Red and yellow circles show two sets of six 100 diffraction spots belonging to different layers.

What is the origin of these extra diffraction spots? One immediate answer is double diffraction, which often occurs in thick two-phase materials when a diffracted beam travelling through a crystal is rediffracted when it passes into a second crystal.[53] Hence, it would be natural to expect that double diffraction induces several extra diffraction spots in MSLs. We, however, argue that the simple double diffraction process cannot account for the extra diffraction spots shown in Fig. 5. Double-diffraction spots occur around each of the primary reflections, including the direct beam spot, by keeping the symmetry of the constituent crystals.[53] This general feature of double diffraction implies that as for twisted hBN layers, two types of primary reflections, each with 6-fold rotational symmetry, should be accompanied by a set of double diffraction spots with the same 6-fold rotational symmetry, as indeed observed in 10° twist angle double-multilayer of ~100-nm-thick hBN sample.[54] Contrary to the above expectation from double diffraction, the extra diffraction spots given in Fig. 5 does not show the 6-fold rotational symmetry except for the nearly commensurate sample at $\theta = 21.6°$. Furthermore, for very thin specimens in which the WPOA holds, the double diffraction does not contribute to the entire diffraction process, as will be discussed below.

For a better understanding of the extra diffraction spots, we analyzed the positions of these spots $k_e$ in reciprocal space. It has been revealed that $k_e$ can be found by connecting the two Bragg spots belonging to the different layers and translating the obtained vectors to the center of the diffraction pattern, as demonstrated in Fig. S6. The relationship can be written as $k_e = G_1 - G_2$, where $G_i$ ($i$ = 1, 2) represents the reciprocal vectors from two different layers. This relationship implies that the interlayer interactions are responsible for the generation of these extra diffraction spots indeed, but the question is, why do particular sets of the reciprocal vectors contribute to the extra diffraction processes? Here, one should remind that under WPOA, the contrast in HRTEM images is proportional to the projected specimen potential, convoluted with the impulse response of the instrument.[51] Note, however, that diffraction peaks due to moiré structures (or double diffraction) cannot be observed when acquired at typical TEM electron energies (60–300 keV) under WPOA, provided that the projected specimen potential consists only of the potentials of individual atoms in the respective layers[55] (see 'Intensity of moiré peaks in diffraction pattern' section in the Supporting Information). We hence assume that the observed extra diffraction spots are originated from a well-defined moiré potential newly created by interlayer coupling. The moiré potential presumably has complicated spatial distributions and periodicity depending on the twist angle. When an electron beam is diffracted by the interlayer moiré potential, the resulting diffraction pattern will not necessarily have 6-fold rotational symmetry except for the commensurate MSLs but should be represented by the combination of a set of $k_e$ vectors with particular frequencies and amplitudes, accounting for the underlying characteristics of the extra diffraction spots with particular frequencies and intensities.

**Construction of the Moiré Potential.** If the assumption mentioned in the previous subsection is valid, the inverse FFT (IFFT) of these extra spots will yield information on the spatial modulation of the moiré potential. It is hence interesting to reconstruct IFFT images from the FFT images by filtering in the frequency domain using the inclusive mask for the extra spots shown in Fig. 5, but excluding all the well-defined Bragg and center spots (for details, see Fig. S7). Figure 7a-d shows the thus obtained IFFT images using the FFT images shown in Fig. 5a-d. We found that regardless of the twisted angles, the IFFT images exhibit a triangular lattice-like pattern. Note also that a similar pattern can be obtained even if the contrast of the respective IFFT images is inverted (Fig. S8). Hence, the resulting images are not artefacts due to IFFT but will represent periodic-like 2D potential functions with comparable positive and negative amplitudes. As for the nearly commensurate cases at $\theta = 21.6°$ and 13.6°, the periodicities of the lattice are estimated to be ~0.6 nm and ~1.0 nm, respectively. These values are comparable to those deduced from Eq. (1) ($D$ = 0.66 nm and 1.08 nm for $\theta = 21.6°$ and 13.6°, respectively). The corresponding line profiles show distinct repeating patterns with positive and negative correlations along the three directions of the triangular lattice (Fig. 7b,d), which extends over the length scale of several to tens of nanometers (see Fig.S9a for a wide-area IFFT image of $\theta = 21.6°$). Thus, the commensurability of the MSLs is almost preserved in the IFFT image. For the samples with $\theta$=24.0° and 19.8°, the translational symmetry is restricted within the range of a few nm (see Fig. S9b for a wide-area IFFT image of $\theta = 24.0°$), which can be viewed as an incommensurate periodicity, in harmony with the incommensurate layered structures.

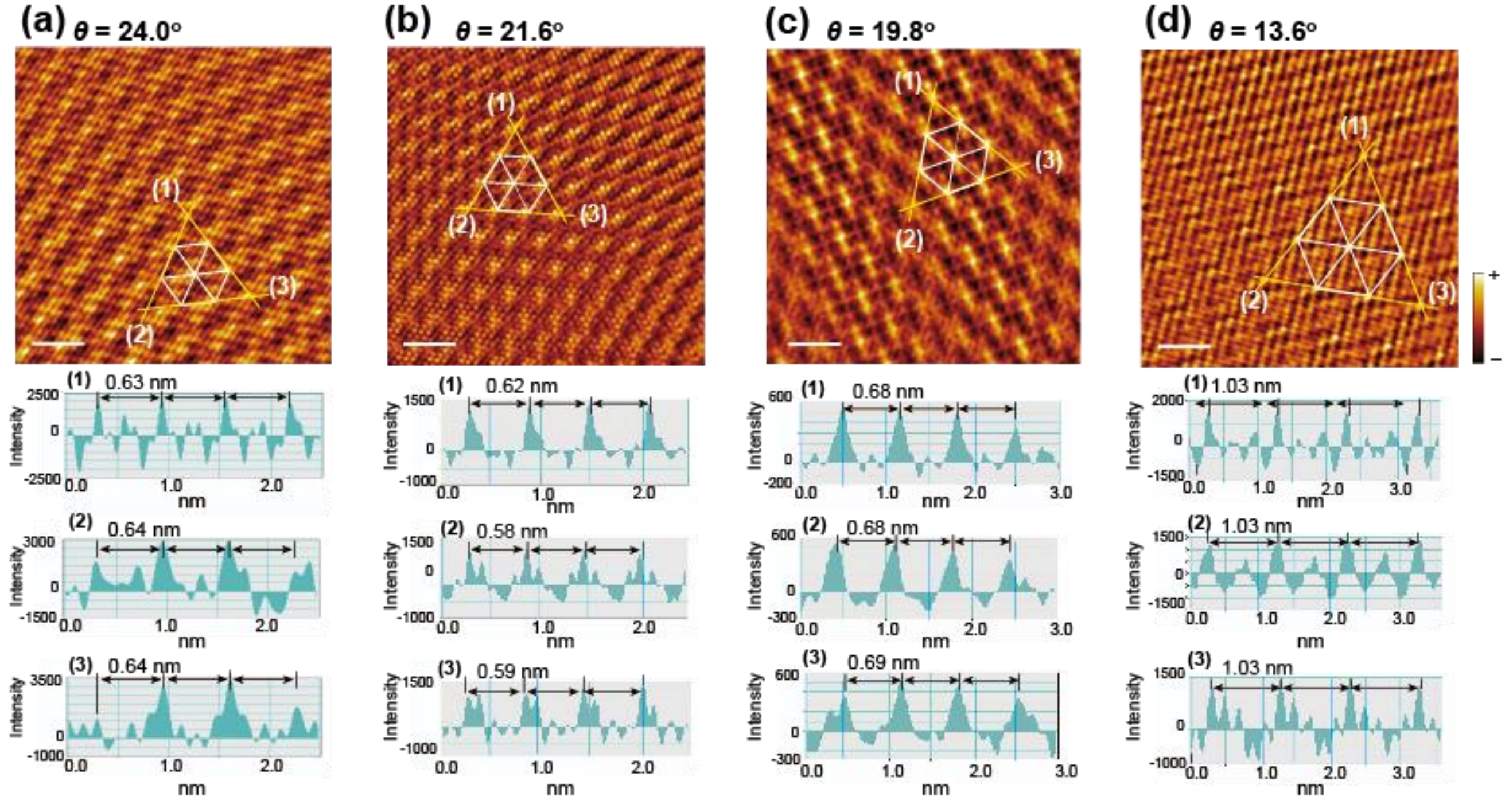

**Figure 7.** Atomic-scale moiré potentials of the twisted hBN layers. (a-d) (top panel) False-color IFFT images obtained by selecting the spots other than the Bragg and center spots shown in Fig. 2a–d, yielding a triangular lattice-like pattern for the samples with twist angles of 24.0° (a), 21.6° (b), 19.8° (c), and 13.6° (d). Scale bar, 1 nm. (lower panels) Line profiles as a function of distance along the three yellow lines (1), (2) and (3) in the corresponding top panel.

**Calculations of electrostatic potential.** The present IFFT images presumably represent the interlayer moiré potential in our incommensurate and nearly commensurate hBN layers. However, additional theoretical support would be necessary to confirm this argument. In theoretical studies of 2D materials, two models are usually employed, i.e., a periodic model and a finite model. The periodic model is based on Bloch's theorem and utilizes periodic boundary conditions to replicate the elementary cell.[14] This approach is in principle applicable only to commensurate systems where the momentum is well defined. In the finite model approach, one performs, for example, quantum chemical calculations using a large enough finite cluster models, whose external dangling bonds are terminated usually by hydrogen atoms.[56,57] Although the results of the finite models may suffer from edge effects,[57] this approach can be applied both to commensurate and incommensurate systems and is useful for calculating electrostatic potential (ESP) in real space. Thus, in this work, we carried out a series of quantum chemical density functional theory (DFT) calculations using sufficiently large clusters of atoms modeling the hBN bilayers with different twist angles as well as the one with the untwisted AA′ stacking configuration ($\theta$ = 0°). Although the DFT calculations were performed on several model clusters with different sizes using various density functionals under the rigid lattice approximation, we only show the results on the $(B_{111}N_{111}H_{42})_2$ cluster at the ωB97XD/6-31G(d) level, as explained in 'Computational details' section in the Supporting Information.

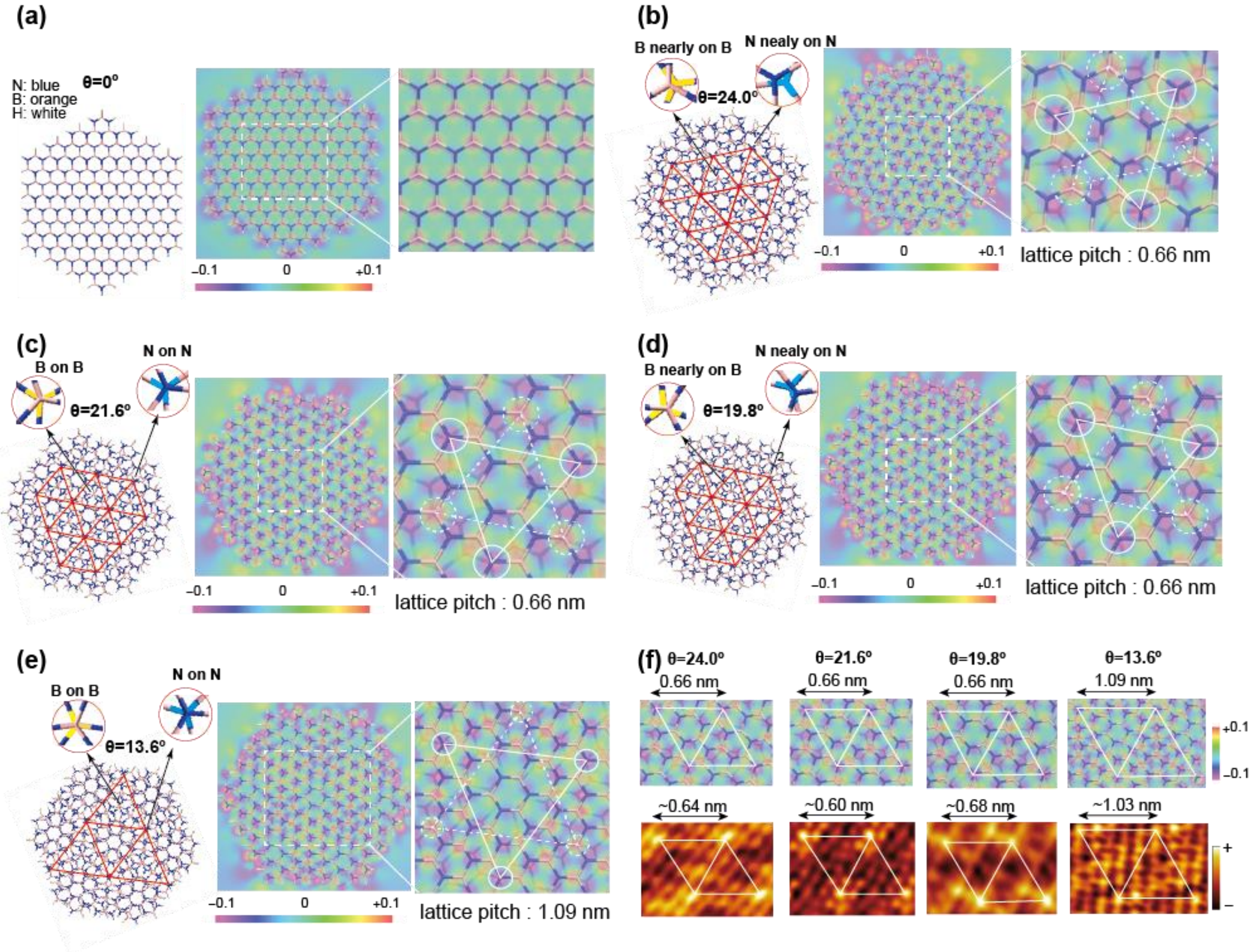

**Figure 8.** DFT cluster calculations of electrostatic potential (ESP). (a-e) Optimized structure (left panel), ESP map (middle panel), and magnified ESP map (right panel) of the $(B_{111}N_{111}N_{42})_2$ clusters with twist angles of (a) 0°, (b) 24.0°, (c) 21.6°, (d) 19.8°, and (e) 13.6°calculated at the ωB97XD/6-31G(d) level. In the left (right) panels, triangular lattice-like patterns created by the interlayer atomic overlap, as shown by the magnified views in the left panels, are indicated by red (white) triangles. The unit of the scale bar is $e$ a.u.$^{-1}$. (f) Comparison between the ESP map (upper panels) and the corresponding IFFT image shown in Fig. 4 (lower panels) with the same twist angle. These images are created by cropping and rotating the original images for the ease of comparison.

Figure 8 shows the resulting optimized structures of the H-terminated clusters with different twist angles, along with the 2D plot of the ESP in a plane bisecting the two hBN layers of the respective clusters. One sees from Fig. 8a that as for the cluster with $\theta$ = 0°, the periodicity of the interlayer ESP matches that of the atomic configuration (for the cross-sectional map perpendicular to the atomic planes, see Fig. S10). The periodic pattern is created by the negative ESP on the N atoms and the positive ESP on the B atoms located exactly above (or below) the N atoms. As the twist angle increases from zero, the anti-aligned AA′ stacking (B on N or N on B) is altered to form various interlayer atomic alignments, such as N on N and B on B configurations, especially for the nearly commensurate clusters (Fig. 8c,e). From the 2D plot of the ESP shown in Fig. 8c,e, one sees that such N/N (B/B) overlaps yield negative (positive) ESP. Consequently, the point where a fully eclipsed arrangement of N/N (B/B) shows the highest local negative (positive) ESP. This leads to the formation of triangular lattice-like patterns with pitches of 0.66 nm and 1.09 nm for $\theta$ = 21.6° and 13.6°, respectively, which are almost equal to those deduced from Eq. (1) ($D$ = 0.66 nm and 1.08 nm for $\theta$ = 21.6° and 13.6°, respectively). Note also that even in the 2D ESP map of incommensurate clusters with θ = 24.0° (Fig. 8b) and 19.8°(Fig. 8d), a similar periodic pattern with a pitch of 0.66 nm is created due to periodicity of N nearly on N (or B nearly on B) configurations. Further worth mentioning is that as shown in Fig. 8f, the triangular lattice-like patterns of the 2D ESP maps are very similar to those found in the IFFT images of the samples with the corresponding twist angles in terms both of the pitch and the pattern. A slight discrepancy between the observed and calculated pitch values may result from the rigid lattice approximation employed in the DFT calculations. This good correspondence confirms our assumption that a series of the IFFT images shown in Fig. 7 represent the interlayer moiré potential created by the local interlayer atomic overlap in the respective hBN bilayers.

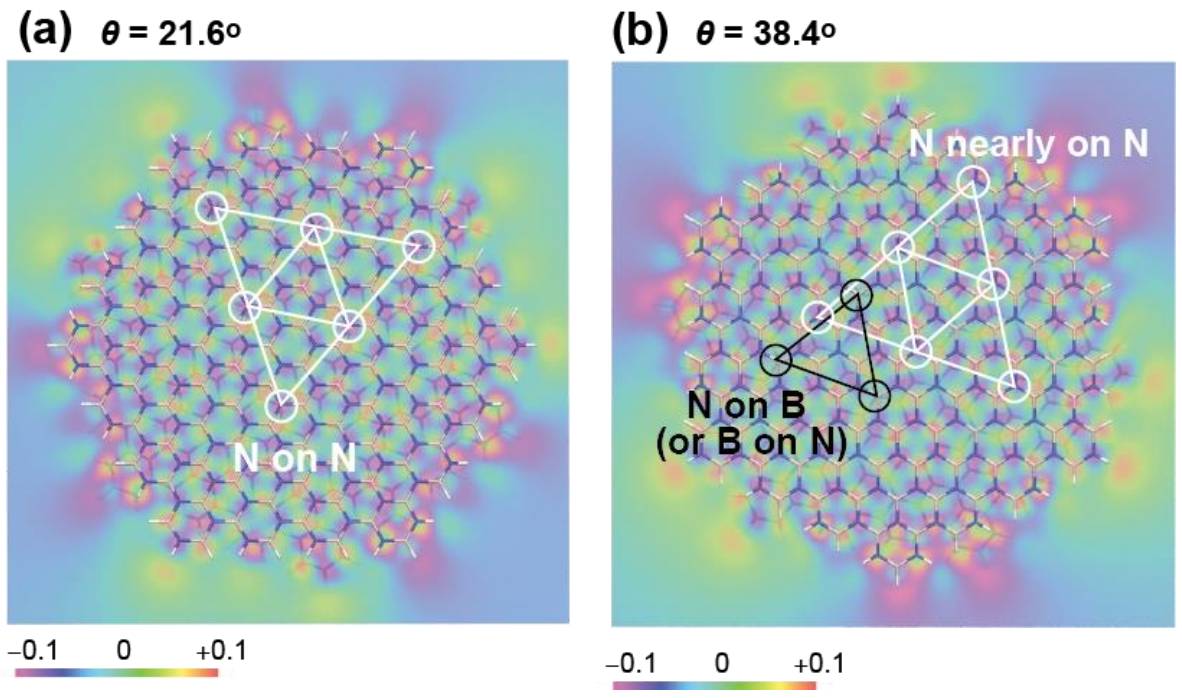

**Figure 9.** ESP map of the $(B_{111}N_{111}N_{42})_2$ clusters with twist angles of (a) 21.6° and (b) 38.4° calculated at the ωB97XD/6-31G(d) level. The unit of the scale bar is $e$ a.u.$^{-1}$.

Thus far, the discussion was given based on the twist angle $\theta$ defined earlier. What happens when we employ an alternative definition of the twist angle, i.e., $\bar{\theta} = 60° - \theta$? To answer the question, we calculated the 2D ESP map of one of the nearly commensurate cluster with $\bar{\theta} = 60° - 21.6° = 38.4°$. In the 38.4° cluster, the aligned N on N (or B on B) stacking points realized in the 21.6° cluster are replaced by the anti-aligned N on B (or B on N) stacking points, yielding locally nearly zero potential regions. Even in the 38.4° cluster, however, there exist a number of N nearly on N (or B nearly on B) configurations, whose periodicity is similar to that of the N on N (or B on B) stacking points in the 21.6° cluster. These N nearly on N (B nearly on B) configurations will yield locally negative (positive) EPS, although their amplitudes are slightly smaller than those for the N exactly on N (B exactly on B) configurations. Consequently, as shown in Fig. 9, the 2D ESP map at 38.4° show a triangular lattice-like pattern with nearly the same periodicity as that at 21.6°, implying that both the ESP maps at 21.6° and 38.4° fit the IFFT image shown in Fig. 7b. Thus, at present we cannot definitely decide which angle, $\theta$ or $\bar{\theta}$, is more consistent with the IFFT image of interest.

Finally, we discuss how the interlayer ESP varies as the interlayer distance $d_{int}$ increases from its optimal value $d_{int(opt)}$. Figure 10 shows a variation in the ESP with interlayer distance obtained for the cluster with $\theta$ = 24.0°, which is characterized by $d_{int(opt)}$ = 0.329 nm. One sees from Fig. 10 that the positive ESP almost disappears for $d_{int}$=0.375 nm, and the interlayer ESP eventually becomes almost structureless for $d_{in}$=0.475 nm. This probably accounts for the reason why in previous HRTEM studies on MSLs, the extra FFT diffraction spots, such as those observed in Fig. 5, have not been reported to occur.[30–35] In the previous HRTEM measurements, it is likely that various adsorbates are trapped between layers during their assembly and sample handling, which prevents the observation of interlayer moiré potential. Thus, as mentioned repeatedly before, the preparation of adsorbate-free samples is a prerequisite to observe moiré potential as FFT diffraction spots. The chemical exfoliation procedure developed in this work is one useful method for obtaining clean hBN nanolayers and hence observing their moiré potential. We are now applying the present method to graphite and transition metal dichalcogenides. Currently, studies are in progress to confirm its applicability to other vdW systems.

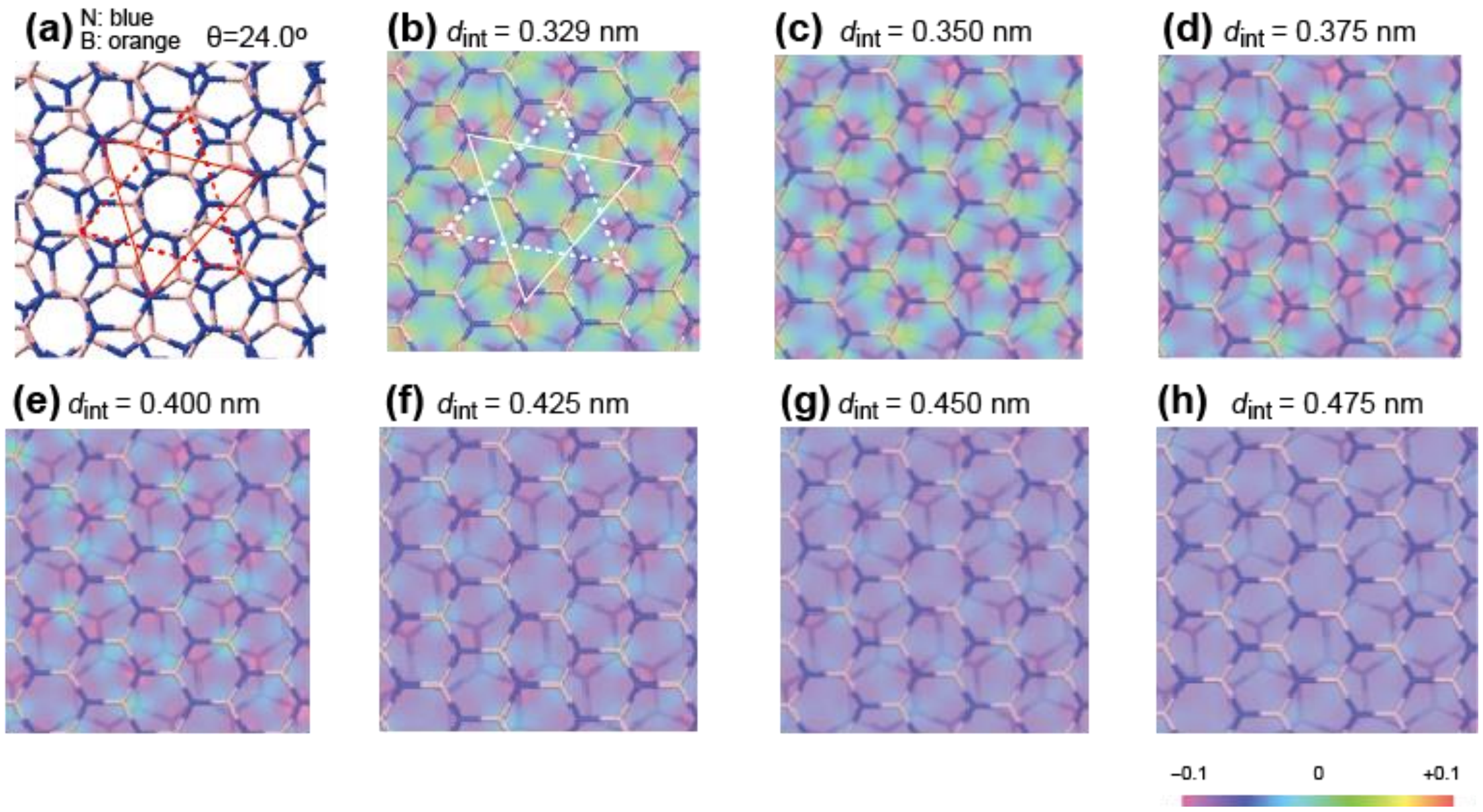

**Figure 10.** Variation in the ESP with interlayer distance. (a) Enlarged representation of the BN bilayer model with a twist angle of 24.0°. (b-h) ESP for the interlayer distance $d_{int}$ of (b) 0.329 nm (optimized value) , (c) 0.350 nm, (d) 0.375 nm, (e) 0.400 nm, (f) 0.425 nm, (g) 0.450 nm, and (h) 0.475 nm. ESP is shown for a plane bisecting the two hBN layers of the respective clusters, and the unit of the scale bar is $e$ a.u.$^{-1}$. A solid (dotted) triangle represents a portion of the triangular-like lattice consisting of N nearly on N (B nearly on B) configurations.

## ■ CONCLUSIONS

We fabricated strain- and contaminant-free hBN layers with different twist angles using the intercalation-based exfoliation technique and observed their HRTEM images. The FFT of the HRTEM images allowed us to recognize moiré diffraction spots, and their IFFT computations provided information about the interlayer moiré potential, as corroborated by the DFT calculations. Also, the DFT calculations highlight the importance of the local interlayer atomic overlap in establishing the moiré potentials both in the commensurate and incommensurate cases. The present results not only pave the way for observing the moiré potentials in twisted hBN layers, but they also expand the application of electron microscopy to visualizing the moiré potential in a variety of MSLs especially with large twist angles.

## ASSOCIATED CONTENT

Supporting Information.

The supporting Information is available free of charge at http://pubs.acs.org. Experimental and computational details, scheme of chemical exfoliation, additional HRTEM and IFFT images, PPA analysis, real and reciprocal representations of hBN bilayers, cross sectional ESP map, and additional DFT results.

## AUTHOR INFORMATION

### Corresponding Author

*Takashi Uchino - *Department of Chemistry, Kobe University, Nada, Kobe 657-8501, Japan*; orchid.org/0000-0002-4899-3078; Email: uchino@kobe-u.ac.jp.

### Author Contributions

The manuscript was written through contributions of all authors. / All authors have given approval to the final version of the manuscript. /

## ACKNOWLEDGMENT

This work was conducted in National Institute for Materials Science (NIMS) and Institute for Molecular Science, supported by "Advanced Research Infrastructure for Materials and Nanotechnology in Japan (ARIM)" of the Ministry of Education, Culture, Sports, Science and Technology (Proposal Numbers JPMXP1223NM0091 and JPMXP1224 MS0004). This work was also supported by NIMS Joint Research Hub Program. The computations were performed in the Research Center for Computational Science, Okazaki, Japan (24-IMS-C194).

## REFERENCES

(1) Park, J. M.; Cao, Y.; Xia, L.-Q.; Sun, S.; Watanabe, K.; Taniguchi, T.; Jarillo-Herrero, P. Robust Superconductivity in Magic-Angle Multilayer Graphene Family. *Nat. Mater.* **2022**, *21* (7), 877–883.

(2) Andrei, E. Y.; Efetov, D. K.; Jarillo-Herrero, P.; MacDonald, A. H.; Mak, K. F.; Senthil, T.; Tutuc, E.; Yazdani A.; Young, A. F. The Marvels of Moiré Materials. *Nat. Rev. Mater.* **2021**, *6* (3), 201–206.

(3) Morales-Durán, N.; Shi, J.; MacDonald, A. H. Fractionalized Electrons in Moiré Materials. *Nat. Rev. Phys.* **2024**, *6* (4), 349–351.

(4) He, F.; Zhou, Y.; Ye, Z.; Cho, S.-H.; Jeong, J.; Meng, X.; Wang, Y. Moiré Patterns in 2D Materials: A Review. *ACS Nano* **2021**, *15* (4), 5944-5958.

(5) Mak K. F.; Shan, J. Semiconductor Moiré Materials. *Nat. Nanotechnol.* **2022**, *17* (7), 686–695.

(6) Nuckolls K. P.; Yazdani, A. A microscopic Perspective on Moiré Materials. *Nat. Rev. Mater.* **2024**, *9* (5), 460–480.

(7) Cao, Y.; Fatemi, V.; Demir, A.; Fang, S.; Tomarken, S. L.; Luo, J. Y.; Sanchez-Yamagishi, J. D.; Watanabe, K.; Taniguchi, T.; Kaxiras, E.; Ashoori, R. C.; Jarillo-Herrero, P. Correlated Insulator Behaviour at Half-Filling in Magic-Angle Graphene Superlattices. *Nature* **2018,** *556*, 80–84.

(8) Cao, Y.; Fatemi, V.; Fang, S.; Watanabe, K.; Taniguchi, T.; Kaxiras E.; Jarillo-Herrero, P. Unconventional Superconductivity in Magic-Angle Graphene Superlattices. *Nature* **2018**, *556*, 43–50.

(9) Saito, Y.; Ge, J.; Watanabe, K.; Taniguchi, T.; Young, A. F. Independent Superconductors and Correlated Insulators in Twisted Bilayer Graphene, *Nat. Phys*. **2020**, *16* (6), 926–930.

(10) Park, H.; Cai, J.; Anderson, E.; Zhang, Y.; Zhu, J.; Liu, X.; Wang, C.; Holtzmann, W.; Hu, C.; Liu, Z.; Taniguchi, T.; Watanabe, K.; Chu, J.-H.; Cao, T.; Fu, L.; Yao, W.; Chang, C.-Z.; Cobden, D.; Xiao D.; Xu, X. Observation of Fractionally Quantized Anomalous Hall Effect. *Nature* **2023**, *622*, 74–79.

(11) Lu, Z.; Han, T.; Yao, Y.; Reddy, A. P.; Yang, J.; Seo, J.; Watanabe, K.; Taniguchi, T.; Fu, L.; Ju, L. Fractional Quantum Anomalous Hall Effect in Multilayer Graphene. *Nature* **2024**, *626*, 759–764.

(12) Dean, C. R.; Wang, L.; Maher, P.; Forsythe, C.; Ghahari, F.; Gao, Y.; Katoch, J.; Ishigami, M.; Moon, P.; Koshino, M.; Taniguchi, T.; Watanabe, K.; Shepard, K. L.; Hone J.; Kim, P. Hofstadter's Butterfly and the Fractal Quantum Hall Effect in Moiré Superlattices. *Nature* **2013**, *497*, 598–602.

(13) Ju, L.; MacDonald, A. H.; Mak, K. F.; Shan J.; Xu, X. The Fractional Quantum Anomalous Hall Effect. *Nat. Rev. Mater.* **2024**, *9* (6), 455–459.

(14) Carr, S.; Fang S.; Kaxiras, E. Electronic-Structure Methods for Twisted Moiré Layers. *Nat. Rev. Mater.* **2020**, *5* (7), 748–763.

(15) Shallcross, S.; Sharma, S.; Kandelaki, E.; Pankratov, O. A. Electronic Structure of Turbostratic Graphene. *Phys. Rev. B* **2010**, *81* (16), 165105.

(16) Carr, S.; Fang, S.; Zhu, Z.; Kaxiras, E. Exact Continuum Model for Low-Energy Electronic States of Twisted Bilayer Graphene. *Phys. Rev. Research* **2019**, *1* (1), 013001.

(17) Mele, E. J. Commensuration and Interlayer Coherence in Twisted Bilayer Graphene. *Phys. Rev. B* **2010**, *81* (16), 161405(R).

(18) Conte, F.; Ninno, D.; Cantele, G. Electronic Properties and Interlayer Coupling of Twisted $MoS_2$/$NbSe_2$ Heterobilayers. *Phys. Rev. B* **2019**, *99* (15), 155429.

(19) Luican, A.; Li, G.; Reina, A.; Kong, J.; Nair, R. R.; Novoselov, K. S.; Geim, A. K.; Andrei, E. Y. Single-Layer Behavior and Its Breakdown in Twisted Graphene Layers. *Phys. Rev. Lett.* **2011**, *106* (12), 126802.

(20) Yao, W.; Wang, E.; Bao, C.; Zhang, Y.; Zhang, K.; Bao, K.; Chan, C. K.; Chen, C.; Avila, J.; Asensio, M. C.; Zhu, J.; Zhou, S. Quasicrystalline 30° Twisted Bilayer Graphene as an Incommensurate Superlattice with Strong Interlayer Coupling. *Proc. Natl. Acad, Sci. USA* **2018**, *115* (27), 6928-6933.

(21) Ahn, S. J.; Moon, P.; Kim, T.-H.; Kim, H.-W.; Shin, H.-C.; Kim, E. H.; Cha, H. W.; Kahng, S.-J.; Kim, P.; Koshino, M.; Son, Y.-W.; Yang, C.-W.; Ahn, J. R. Dirac Electrons in a Dodecagonal Graphene Quasicrystal. *Science* **2018**, *361* (6404), 782-786.

(22) Pezzini, S.; Mišeikis, V.; Piccinini, G.; Forti, S.; Pace, S.; Engelke, R.; Rossella, F.; Watanabe, K.; Taniguchi, T.; Kim, P.; Coletti, C. 30°-Twisted Bilayer Graphene Quasicrystals from Chemical Vapor Deposition, *Nano Lett.* **2020**, *20* (5), 3313-3319.

(23) Li, Y.; Zhang, F.; Ha, V.-A.; Lin, Y.-C.; Dong, C.; Gao, Q.; Liu, Z.; Liu, X.; Ryu, S. H.; Kim, H.; Jozwiak, C.; Bostwick, A.; Watanabe, K.; Taniguchi, T.; Kousa, B.; Li, X.; Rotenberg, E.; Khalaf, E.; Robinson, J. A.; Giustino F.; Shih, C.-K. Tuning Commensurability in Twisted van der Waals Bilayers. *Nature* **2024**, *625*, 494–499.

(24) Sponza, L.; Vu, V. B.; Richaud, E. S.; Amara, H.; Latil, S. Emergence of Flat Bands in the Quasicrystal Limit of Boron Nitride Twisted Bilayers. *Phys. Rev. B* **2024**, *109* (16), L161403.

(25) Agarwal, H.; Nowakowski, K.; Forrer, A.; Principi, A.; Bertini, R.; Batlle-Porro, S.; Reserbat-Plantey, A.; Prasad, P.; Vistoli, L.; Watanabe, K.; Taniguchi, T.; Bachtold, A.; Scalari, G.; Kumar R. K.; Koppens, F. H. L. Ultra-Broadband Photoconductivity in Twisted Graphene Heterostructures with Large Responsivity. *Nat. Photon.* **2023**, *17* (9), 1047–1053.

(26) Bucko, J.; Herman, F. Large Twisting Angles in Bilayer Graphene Moiré Quantum Dot Structures. *Phys. Rev. B* **2021**, *103* (7), 075116.

(27) Deng B.; Xia, F. Large-Angle Twist Effect. *Nat. Photon.* **2023**, *17*, 1021–1022.

(28) Woods, C. R.; Britnell, L.; Eckmann, A.; Ma, R. S.; Lu, J. C.; Guo, H. M.; Lin, X.; Yu, G. L.; Cao, Y.; Gorbachev, R. V.; Kretinin, A. V.; Park, J.;. Ponomarenko, L. A.; Katsnelson, M. I.; Gornostyrev, Yu. N.; Watanabe, K.; Taniguchi, T.; Casiraghi, C.; Gao, H-J., Geim A. K.; Novoselov, K. S. Commensurate–Incommensurate Transition in Graphene on Hexagonal Boron Nitride. *Nat. Phys.* **2014**, *10* (4), 451–456.

(29) Deng, B.; Wang, B.; Li, N.; Li, R.; Wang, Y.; Tang, J.; Fu, Q.; Tian, Z.; Gao, P.; Xue, J.; Peng, H. Interlayer Decoupling in 30° Twisted Bilayer Graphene Quasicrystal. *ACS Nano* **2020**, *14* (2), 1656-1664.

(30) Robertson, A. W.; Bachmatiu, A.; Wu, Y. A.; Schäffel, F.; Büchner, B.; Rümmeli, M. H.; Warner, J. H. Structural Distortions in Few-Layer Graphene. *ACS Nano* **2011**, *5* (12), 9984–9991.

(31) Zan , R.; Bangert, U.; Ramasse, Q.; Novoselov, K. S. Imaging of Bernal Stacked and Misoriented Graphene and Boron Nitride: Experiment and Simulation. *J. Microscopy* **2011**, *244* (2), 152-158.

(32) Yuan, S.; Linas, S.; Journet, C.; Steyer, P.; Garnier, V.; Bonnefont, G.; Brioude, A.; Toury, B. Pure & Crystallized 2D Boron Nitride Sheets Synthesized via a Novel Process Coupling both PDCs and SPS methods. *Sci. Rep.* **2016**, *6*, 20388.

(33) Warner, J. H.; Rümmeli, M. H.; Gemming, T.; Büchner, B.; Andrew, G.; Briggs, D. Direct Imaging of Rotational Stacking Faults in Few Layer Graphene. *Nano Lett*. **2009**, *9* (1), 102-106.

(34) Bachmatiuk, A.; Zhao, J.; Gorantla, S. M.; Martinez, I. G. G.; Wiedermann, J.; Lee, C.; Eckert, J.; Rummeli, M. H. Low Voltage Transmission Electron Microscopy of Graphene. *Small* **2015**, *11*(5), 515-542.

(35) Yamazaki, K.; Maehara, Y.; Gohara, K. Characterization of TEM Moiré Patterns Originating from Two Monolayer Graphenes

Grown on the Front and Back Sides of a Copper Substrate by CVD Method. *J. Phys. Soc. Jpn.* **2018**, *87* (6), 061011.
(36) Haigh, S. J.; Gholinia, A.; Jalil, R.; Romani, S.; Britnell, L.; Elias, D. C.; Novoselov, K. S.; Ponomarenko, L. A.; Geim A. K.; Gorbachev, R. Cross-Sectional Imaging of Individual Layers and Buried Interfaces of Graphene-Based Heterostructures and Superlattices. *Nat. Mater.* **2012**, *11* (7), 764–767.
(37) Gasparutti, I.; Song, S. H.; Neumann, M.; Wei, X.; Watanabe, K.; Taniguchi, T.; Lee, Y. H. How Clean Is Clean? Recipes for van der Waals Heterostructure Cleanliness Assessment. *ACS Appl. Mater. Interfaces* **2020**, *12* (6), 7701–7709.
(38) Rosenberger, M. R.; Chuang, H.-J.; McCreary, K. M.; Hanbicki, A. T.; Sivaram, S. V.; Jonker, B. T. Nano-“Squeegee” for the Creation of Clean 2D Material Interfaces. *ACS Appl. Mater. Interfaces* **2018**, *10* (12), 10379–10387.
(39) Cai L.; Yu, G. Fabrication Strategies of Twisted Bilayer Graphenes and Their Unique Properties. *Adv. Mater.* **2021**, *33* (13), 2004974.
(40) Liao, M.; Wei, Z.; Du, L.; Wang, Q.; Tang, J.; Yu, H.; Wu, F.; Zhao, J.; Xu, X.; Han, B.; Liu, K.; Gao, P.; Polcar, T.; Sun, Z.; Shi, D.; Yang R.; Zhang, G. Precise Control of the Interlayer Twist Angle in Large Scale $MoS_2$ Homostructures. *Nat. Commun*. **2020**, *11*, 2153.
(41) Yin, J.; Li, J.; Hang, Y.; Yu, J.; Tai,G.; Li, X.; Zhang, Z.; Guo, W. Boron Nitride Nanostructures: Fabrication, Functionalization and Applications. *Small* **2016**, *12* (22), 2942–2968.
(42) Kim, D. S.; Dominguez, R. C.; Mayorga-Luna, R.; Ye, D.; Embley, J.; Tan, T.; Ni, Y.; Liu, Z.; Ford, M.; Gao, F. Y.; Arash, S.; Watanabe, K.; Taniguchi, T.; Kim, S.; Shih, C.-K.; Lai, K.; Yao, W.; Yang, L.; Li, X.; Miyahara, Y. Electrostatic Moiré Potential from Twisted Hexagonal Boron Nitride Layers. *Nat. Mater.* **2024**, *23* (8), 65–70.
(43) Yang, R.; Fan, Y.; Mei, L.; Shin, H. S.; Voiry, D.; Lu, Q.; Li, J.; Zeng, Z. Synthesis of Atomically Thin Sheets by the Intercalation-Based Exfoliation of Layered Materials. *Nat. Synth.* **2023**, *2* (2), 101–118.
(44) Kovtyukhova, N. I.; Wang, Y.; Lv, R.; Terrones, M.; Crespi, V. H.; Mallouk, T. E. Reversible Intercalation of Hexagonal Boron Nitride with Brønsted Acids. *J. Am. Chem. Soc.* **2013**, *135* (22), 8372.
(45) Tsujimura, T.; Uchino, T. Oriented Crystallization of Ammonium Sulfate from Hexagonal Boron Nitride/Sulfuric Acid Intercalation Compounds. *ACS Omega* **2021**, *6* (9) , 6482-6487.
(46) Galindo, P. L.; Kret, S.; Sanchez, A.M.; Laval, J.-Y.; Yáñez, A.; Pizarro, J.; Guerrero, E.; Ben, T.; Molina, S. I. The Peak Pairs Algorithm for Strain Mapping from HRTEM Images. *Ultramicroscopy* **2007**, *107* (12), 1186-1193.
(47) Frisch, M. J.; Trucks, G. W.; Schlegel, H. B.; Scuseria, G. E.; Robb, M. A.; Cheeseman, J. R.; Scalmani, G.; Barone, V.; Petersson, G. A.; Nakatsuji, H.; Li, X.; Caricato, M.; Marenich, A. V.; Bloino, J.; Janesko, B. G.; Gomperts, R.; Mennucci, B.; Hratchian, H. P.; Ortiz, J. V.; Izmaylov, A. F.; Sonnenberg, J. L.; Williams-Young, D.; Ding, F.; Lipparini, F.; Egidi, F.; Goings, J.; Peng, B.; Petrone, A.; Henderson, T.; Ranasinghe, D.; Zakrzewski, V. G.; Gao, J.; Rega, N.; Zheng, G.; Liang, W.; Hada, M.; Ehara, M.; Toyota, K.; Fukuda, R.; Hasegawa, J.; Ishida, M.; Nakajima, T.; Honda, Y.; Kitao, O.; Nakai, H.; Vreven, T.; Throssell, K.; Montgomery, J. A., Jr.; Peralta, J. E.; Ogliaro, F.; Bearpark, M. J.; Heyd, J. J.; Brothers, E. N.; Kudin, K. N.; Staroverov, V. N.; Keith, T. A.; Kobayashi, R.; Normand, J.; Raghavachari, K.; Rendell, A. P.; Burant, J. C.; Iyengar, S. S.; Tomasi, J.; Cossi, M.; Millam, J. M.; Klene, M.; Adamo, C.; Cammi, R.; Ochterski, J. W.; Martin, R. L.; Morokuma, K.; Farkas, O.; Foresman, J. B.; Fox, D. J. *Gaussian 16*, Revision C.02; Gaussian, Inc.: Wallingford, CT, 2019.
(48) Krivanek, O. L.; Chisholm, M. F.; Nicolosi, V.; Pennycook, T. J.; Corbin, G. J.; Dellby, N.; Murfitt, M. F.; Own, C. S.; Szilagyi, Z. S.; Oxley, M. P.; Pantelides, S. T.; Pennycook, S. J. Atom-by-Atom Structural and Chemical Analysis by Annular Dark-Field Electron Microscopy. *Nature* **2010**, *464*, 571–574.
(49) Odlyzko, M. L.; Mkhoyan, K. A. Identifying Hexagonal Boron Nitride Monolayers by Transmission Electron Microscopy. *Microsc. Microanal.* **2012**, *18*, 558-567.
(50) Seto, Y.; Ohtsuka, M. ReciPro: Free and Open-Source Multipurpose Crystallographic Software Integrating a Crystal Model Database and Viewer, Diffraction and Microscopy Simulators, and Diffraction Data Analysis Tools. *J. Appl. Cryst*. **2022**, *55*(2), 397–410.
(51) Spence, J. C. H. *High-Resolution Electron Microscopy*, 4th ed.; Oxford University Press: New York, NY, 2013.
(52) Gass, M. H.; Bangert, U.; Bleloch, A. L.; Wang, P.; Nair, R. R.; Geim, A. K. Free-Standing Graphene at Atomic Resolution. *Nat. Nanotechnol*. **2008**, *3* (9), 676–681.
(53) Williams, D. B; Carter, C. B. Transmission *Electron Microscopy: A Textbook for Materials Science*, 2nd ed.; Springer: New York, NY, 2009.
(54) Lee, H. Y.; Al Ezzi, M. M. ; Raghuvanshi, N. ; Chung, J. Y.; Watanabe, K.; Taniguchi, T.; Garaj, S.; Adam, S.; Gradečak, S. Tunable Optical Properties of Thin Films Controlled by the Interface Twist Angle, *Nano Lett*. **2021**, *21*(7), 2832–2839.
(55) Latychevskaia, T.; Escher, C.; Fink, H.-W. Moiré Structures in Twisted Bilayer Graphene Studied by Transmission Electron Microscopy. *Ultramicroscopy* **2019**, *197*, 46-52.
(56) Mendelson, N.; Chugh, D.; Reimers, J. R.; Cheng, T. S.; Gottscholl, A.; Long, H.; Mellor, C. J.; Zettl, A.; Dyakonov, V.; Beton, P. H.; Novikov, S. V.; Jagadish, C.; Tan, H. H.; Ford, M. J.; Toth, M.; Bradac, C.; Aharonovich, I. Identifying Carbon as the Source of Visible Single-Photon Emission from Hexagonal Boron Nitride. *Nat. Mater.* **2021**, *20* (11), 321–328.
(57) Karamanis, P.; Otero, N.; Pouchan, C. Electric Property Variations in Nanosized Hexagonal Boron Nitride/Graphene Hybrids. *J. Phys. Chem. C* **2015**, *119* (21), 11872–11885.

Insert Table of Contents artwork here

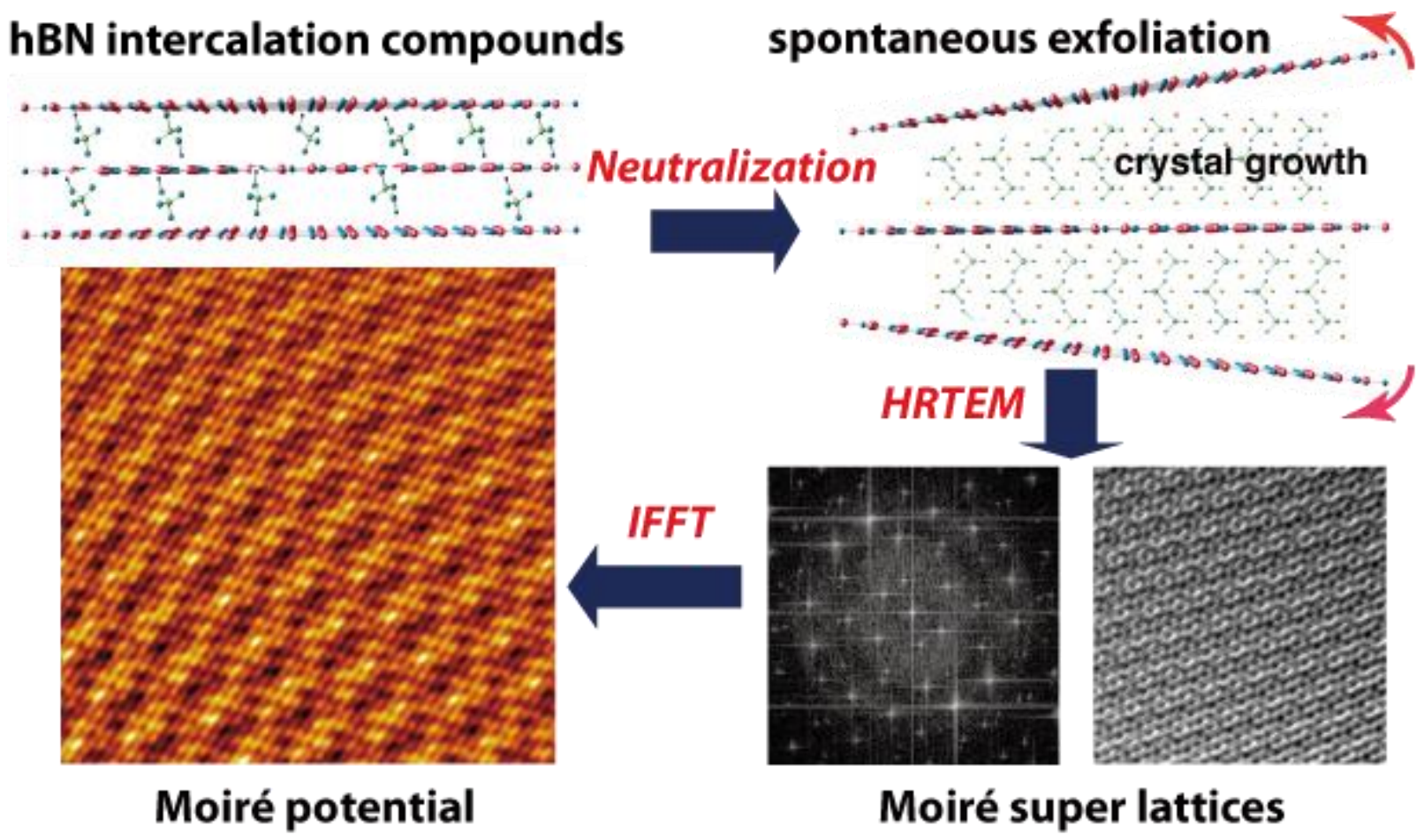

**Supporting Information:**

# Atomic-Scale Observation of Moiré potential in Twisted Hexagonal Boron Nitride Layers by Electron Microscopy

Rina Mishima,[†] Takuro Nagai,[§] Hiroyo Segawa,[ǁ] Masahiro Ehara,[‡] and Takashi Uchino[†]*

[†]Department of Chemistry, Graduate School of Science, Kobe University, Nada, Kobe 657-8501, Japan

[§]Research Network and Facility Services Division, National Institute for Materials Science, Tsukuba, Ibaraki 305-0044, Japan

[ǁ]Research Center for Electronic and Optical Materials, National Institute for Materials Science, Tsukuba, Ibaraki 305-0044, Japan

[‡]Research Center for Computational Science, Institute for Molecular Science, Okazaki, Aichi, 444-8585, Japan

*Corresponding author: uchino@kobe-u.ac.jp

## Table of Contents

## S1. Sample preparation

Boron Nitride was purchased from Kojundo Chemical Lab. Co., Ltd. (purity ~99%, particle size ~10 μm) without any additional treatment. We confirmed that the sample powders show Bragg X-ray diffraction peaks attributed only to the hBN phase. Sulfuric acid (95 %) was obtained from Wako Pure Chemical Industries, Ltd. and was used as purchased. We prepared h-BN/$H_2SO_4$ intercalation compounds according to the scheme shown schematically in Fig. S1. A 30 mg amount of h-BN was added to 0.5 mL of sulfuric acid in a 50 mL Teflon-lined stainless steel autoclave. The autoclave was heated at 200 °C for 24 h, and then naturally cooled to room temperature, producing a slightly brownish slurry. The slurry was transferred into a test tube and was neutralized with an aqueous solution of NaHCO3, resulting in an opaque solution. During neutralization, sodium sulfate salt ($Na_2SO_4$) is expected to be formed in between the hBN layers, leading to spontaneous exfoliation of hBN layers The solution was centrifuged at 15000 rpm for 45 min to remove insoluble materials, and then the supernatant was transferred to a new test tube. The thus obtained supernatant is an aqueous solution containing $Na_2SO_4$ and exfoliated hBN nanosheets. A solvent extraction method was utilized to isolate the exfoliated hBN nanosheets from the supernatant solution. That is, the solution was mixed with an equal volume of organic solvent in a separatory funnel, leading to migration of hBN nanosheets into the organic solvent. We found that 1-pentanol is the most suitable solvent for the present purpose. Although the extracted 1-pentanol solvent was transparent, a colloidal dispersion of the exfoliated hBN nanosheets was confirmed by the Tyndall effect (Fig. S1).

For the TEM observations, the extracts of a volume of ~20 mL were concentrated to a volume of about 0.7 mL, then *ethanol of* 20 mL was added. This solvent exchange is useful to minimize the amount of hydrocarbon adsorbates on the sample. Then, 5 μL of the suspension was collected with a pipette and drop casted onto a TEM microgrid. The TEM grid was dried in vacuum using a dry pumping station for ~2 h to remove solvent.

## S2. TEM observations

Atomic-resolution HRTEM and STEM observations were carried out on a JEOL JEM-ARM300F instrument equipped with a cold field emission electron gun and a spherical aberration ($C_s$) corrector at an acceleration voltage of 80 kV, under $1 \times 10^{-5}$ Pa in the specimen column. Typically, we adjusted the $C_s$ value to 1 μm. All HRTEM images were recorded with an exposure time of 4 s on a Complementary Metal-Oxide-Semiconductor (CMOS) camera (Gatan OneView, 4096 × 4096 pixels). In ADF STEM imaging, the probe size was 0.07 nm, the convergence angle was 32 mrad, and the inner and outer collection angles were 45 and 180 mrad, respectively. Although the operated voltage appears to be higher than the knock-on radiation damage threshold reported for h-BN (78 kV),[58] we did not observe any radiation damage during the TEM observations. Image processing including the fast Fourier transformation (FFT) and inverse FFT (IFFT) was carried out by Gatan Digital Micrograph.

We analysed strain of the exfoliated hBN layers using the Peak Pairs Analysis (PPA) software package (HREM research) for Gatan Digital Micrograph.[46] For the analysis, we employed low-pass filtered HRTEM images of the non-twisted hBN layers. The peak positions were determined on the filtered image, and the relative displacement fields ($u_x$, $u_y$) of the measured lattice with respect to a reference basis vector were calculated. The components of strain tensor $\varepsilon_{xx}$, $\varepsilon_{xy}$, $\varepsilon_{yy}$, mean dilatation $\Delta_{xy}$, rotation $\omega_{xy}$ (in radians) are defined as follows:

$$\varepsilon_{xx} = \frac{\partial u_x}{\partial x},\ \varepsilon_{xy} = \frac{1}{2}\left(\frac{\partial u_x}{\partial y} + \frac{\partial u_y}{\partial x}\right),\ \varepsilon_{yy} = \frac{\partial u_y}{\partial y},$$

$$\Delta_{xy} = \frac{1}{2}(\varepsilon_{xx} + \varepsilon_{yy}),\ \omega_{xy} = \frac{1}{2}\left(\frac{\partial u_y}{\partial x} - \frac{\partial u_x}{\partial y}\right).$$

**S3. Intensity of moiré peaks in diffraction pattern**

We explain how the moiré peaks in the diffraction pattern of twisted bilayers are generated. First, we consider the electron transmission through a non-twisted 2D crystal. The total transmission function $t(x, y)$ of the sample is described by:

$$t(x, y) = \exp[-i\Phi(x, y)] = \exp[-i\sigma V(x, y)] \quad \text{(S1)}$$

where $\Phi(x, y)$ is the sample phase shift, $\sigma$ is the interaction parameter, $V(x, y)$ is the projected potential of the entire sample. The interaction parameter $\sigma$ is given by $\sigma = \frac{2\pi m e \lambda}{h^2}$, where $m$ is the relativistic mass of the electron, $e$ is the elementary charge, $\lambda$ is the wavelength of the electrons, and $h$ is the Planck's constant. Eq. (S1) can further be approximated by the first order Taylor series as

$$t(x, y) \approx 1 - i\sigma V(x, y) \text{ (S2)}$$

under the weak-phase object approximation (WPOA),[51] which is applicable to very thin crystals of light atoms, as in the case of hBN nanolayers.

Then we move on to the case of twisted bilayers. For the time being, we do not take into account the effect of the interlayer moiré potential in the electron transmission process. When neglecting the diffraction effects due to propagation between the two layers, the transmission function of the twisted bilayers can be written as under WPOA:

$$t(x, y) = [1 - i\sigma V_t(x, y)][1 - i\sigma V_b(x, y)]$$
$$= 1 - i\sigma V_t(x, y) - i\sigma V_b(x, y) - \sigma^2 V_t(x, y) V_b(x, y) \text{ (S3)}$$

where $V_t(x, y)$ and $V_b(x, y)$ are the projected potentials of the top and the bottom layers, respectively. The last term describes the overlap of the top- and bottom-layer potentials, which can, in principle, contribute to the formation of moiré peaks in the diffraction pattern. In high-energy (20 – 300 keV) electron microscopy, however, the related moiré peaks are not observed because of the following reasons.[55]

From Eq. (S3), one sees that the second and third terms, which describes the intensity of the main peaks, are proportional to $\sigma$, whereas the last term to $\sigma^2$. Note, however, that for typical TEM electron energies (20 – 300 keV), the value of $\sigma$ is relatively small, e.g., $\sigma = 0.81$ (keV · Å)$^{-1}$ at 100 keV. Consequently, the last term proportional to $\sigma^2$ becomes negligibly small, as compared to the second and third terms, resulting in the absence of moiré peaks in conventional TEM mode. When the electron energy is reduced below ~100 eV, moiré peaks should be visible as $\sigma$ becomes relatively high,[55] e.g., $\sigma = 25.61$ (keV · Å)$^{-1}$ for 100 eV. However, atomic-scale imaging is not possible for such low electron energies.

Finally, we take account of the interlayer moiré potential in the electron transmission process. We assume that the interlayer moiré potential is created in between the top and bottom layers as a result of the interlayer coupling and also that this potential explicitly contributes to the phase shift. The resulting transmission function of the twisted bilayers can be described as:

$$t(x, y) = [1 - i\sigma V_t(x, y)][1 - i\sigma V_b(x, y)][1 - i\sigma V_{int}(x, y)]$$

$$= 1 - i\sigma V_t(x,y) - i\sigma V_b(x,y) - i\sigma V_{int}(x,y) \ldots \quad \text{(S4)}$$

where $V_{int}(x,y)$ describes the spatial modulation of the moiré potential in the interlayer region. In this case, the last term in Eq. (S4) is proportional to $\sigma$, and the intensity of the resulting peaks due to the moiré potential is expected to be comparable to that of the main peaks even for typical TEM electron energies. This argument provides a reasonable explanation for the occurrence of the extra diffraction spots shown in Fig. 5. Hence, we suggest that the absence of the moiré peaks in previous TEM studies on MSLs[30–35] should not be ascribed to the high TEM electron energies, but rather to the absence of interlayer moiré potential due to the strains and contaminants remained in the samples investigated. Considering that the effect of interlayer decoupling becomes more significant with increasing twist angle,[59,60] we would assert that cleanliness of the sample is a matter of great importance in exploring the moiré potential especially for the samples with larger twist angles (see also Fig. 10).

## S4. Computational details

All quantum chemical calculations were performed by density functional theory (DFT) methods implemented in the Gaussian 16 suite of programs[47] using the 6-31G(d) basis set. The suitable selection of functional, and particularly the treatment of long-range exchange interactions in the functional, is crucial for correctly describing vdW interaction.[61] To provide comparison, two functionals were used: one is the B3LYP with the dispersion corrections for the non-bonding vdW interactions (B3LYP-D3),[62,63] and other is a long-range corrected hybrid density functional (ωB97XD)[64] with a parameterized classical dispersion term.

In this work, hydrogen terminated clusters were used to model the local structure of twisted hBN bilayers with four different twist angles ($\theta$ = 13.6°, 19.8°, 21.6° and 24.0°). For comparison, we also employed a cluster with zero twist angle, modeling the local structure of pristine hBN bilayers with the AA′ high-symmetry stacking. To bear out the electronic structure characteristics of the systems considered, we carefully checked how the size and edge shape of the clusters affect or do not affect the calculated results. For this purpose, we employed three types of clusters. Two of them are $(B_{75}N_{75}H_{30})_2$ and $(B_{147}N_{147}H_{42})_2$ bearing zigzag edges, and the other one is $(B_{111}N_{111}H_{42})_2$ featuring armchair edges (Fig. S11). Considering that the twist angles examined in this study are larger than 10° and circumventing possible edge effects of the clusters, we employed the rigid lattice assumption in the structural optimization process; that is, we assumed that each BN layer is completely flat and consists of ideal BN hexagons, in which all the B–N distances are identical and all the B–N–B and N–B–N bond angles are 60°. As a result of the above structural constraints, the optimized structural parameters are the B–N distance ($d$B–N), the interlayer distance between the two BN layers ($d$int), and the terminal B-H and N-H bond distances.

Since our main concern is how the optimized structure, especially the B-N bond distance $d_{\text{B–N}}$ and the interlayer distance $d_{\text{int}}$, are influenced by the size of the clusters and/or the type of edges (zigzag or armchair), we obtained the optimized values of $d_{\text{B–N}}$ and $d_{\text{int}}$ for the above three types of clusters with twist angles of $\theta$ = 0° and 24.0° at the B3LYP-D3/6-31G(d) level. Figure S11 summarizes the results of geometry optimization for the above three model clusters, showing that the optimized structural parameters depend hardly on the type of the clusters. We also found that irrespective of the types of the clusters, the optimized values of $d_{\text{B–N}}$ and $d_{\text{int}}$ obtained for the cluster with $\theta$ = 0° lie in the range 0.1445–0.1446 nm and 0.3332–0.3339 nm, respectively, which are in good agreement with the experimental values of bulk hBN ($d_{\text{B–N}}$=0.145 nm, $d_{\text{int}}$= 0.333 nm),[65] The good agreement between the calculated and observed bulk values of $d$int is likely to be fortuitous because the reported $d$int value for hBN multilayers is slightly shorter ($d_{\text{int}}$= 0.325 nm)[66] than the bulk one (see also Fig. 4b). When the value of $\theta$ increases to 24.0°, the optimized values of $d_{\text{B–N}}$ remain constant, whereas those of $d_{\text{int}}$ show a slight increase for all the clusters employed. This implies that an increase in $\theta$ leads to a decrease in the interlayer interaction. For the calculation of the electrostatic potential (ESP), we employ the $(B_{111}N_{111}H_{42})_2$ cluster with armchair edges because the $(B_{147}N_{147}H_{42})_2$ cluster is too large to perform such calculations within a reasonable computational cost.

The ESP at position $\boldsymbol{r}$ ($\phi_{\text{ESP}}(\boldsymbol{r})$) is given as a sum of contributions from the nuclei and the electronic wave function $\Psi$ given by,

$$\phi_{\mathrm{ESP}}(\boldsymbol{r}) = \sum_{\mathrm{A}}^{\mathrm{nuclei}} \frac{Z_{\mathrm{A}}}{|\boldsymbol{r} - \boldsymbol{R}_{\mathrm{A}}|} - \int \frac{|\Psi(\boldsymbol{r}')|^2}{|\boldsymbol{r} - \boldsymbol{r}'|} d\boldsymbol{r}', \quad \text{(S5)}$$

where A represents a nucleus, and $Z_{\mathrm{A}}$ and $\boldsymbol{R}_{\mathrm{A}}$ are the charge and position of the nucleus A, respectively. The ESP is generally insensitive to the level of sophistication, e.g., the size of the basis set and the level of electron correlation, since it depends directly on the electron density ($\rho = |\Psi|^2$).[67] 2D ESP was rendered by Visual Molecular Dynamics (VMD)[68] and/or Gaussview[69] software packages.

The effect of density functional on the optimized structure and ESP is also of interest. Figure S12 shows the optimized values of *d*B–N and *d*int obtained for the (B111N111H42)2 cluster with $\theta = 0°$, 13.6°, 19.8°, 21,6° and 24.0° calculated at the B3LYP-D3/6-31G(d) and ωB97XD/6-31G(d) levels. The optimized values of $d_{\mathrm{B–N}}$ and $d_{\mathrm{int}}$ at the ωB97XD/6-31G(d) level yield smaller values than those at the B3LYP-D3/6-31G(d) level irrespective of the twist angle; that is, the value of *d*int ($d_{\mathrm{B–N}}$) at the ωB97XD/6-31G(d) level is ~0.01 nm (~0.003 nm) smaller than that at the B3LYP-D3/6-31G(d) level. Consequently, as for the $(B_{111}N_{111}H_{42})_2$ cluster with $\theta = 0°$, the ωB97XD/6-31G(d) level yields the *d*int value of 0.3241 nm, which is quite comparable to that of hBN multilayers ($d_{\mathrm{int}}$ = 0.325 nm).[66] This suggests that the long-range electron-electron exchange interactions are reasonably incorporated in the ωB97XD functional. The interlayer ESPs calculated at the B3LYP-D3/6-31G(d) and ωB97XD/6-31G(d) levels for the $(B_{111}N_{111}H_{42})_2$ cluster with $\theta = 21.6°$ and 13.6° are given in Fig. S13. We did not find any critical differences between the ESPs at the B3LYP-D3/6-31G(d) and ωB97XD/6-31G(d) levels, although these two levels of calculations result in somewhat different values of $d_{\mathrm{int}}$. This implies that the underlying features of the interlayer ESP are insensitive to the DFT functional. Hence, in the main text, we only show the ESP map for the $(B_{111}N_{111}H_{42})_2$ cluster calculated at the ωB97XD/6-31G(d) level.

## S5. Supporting figures

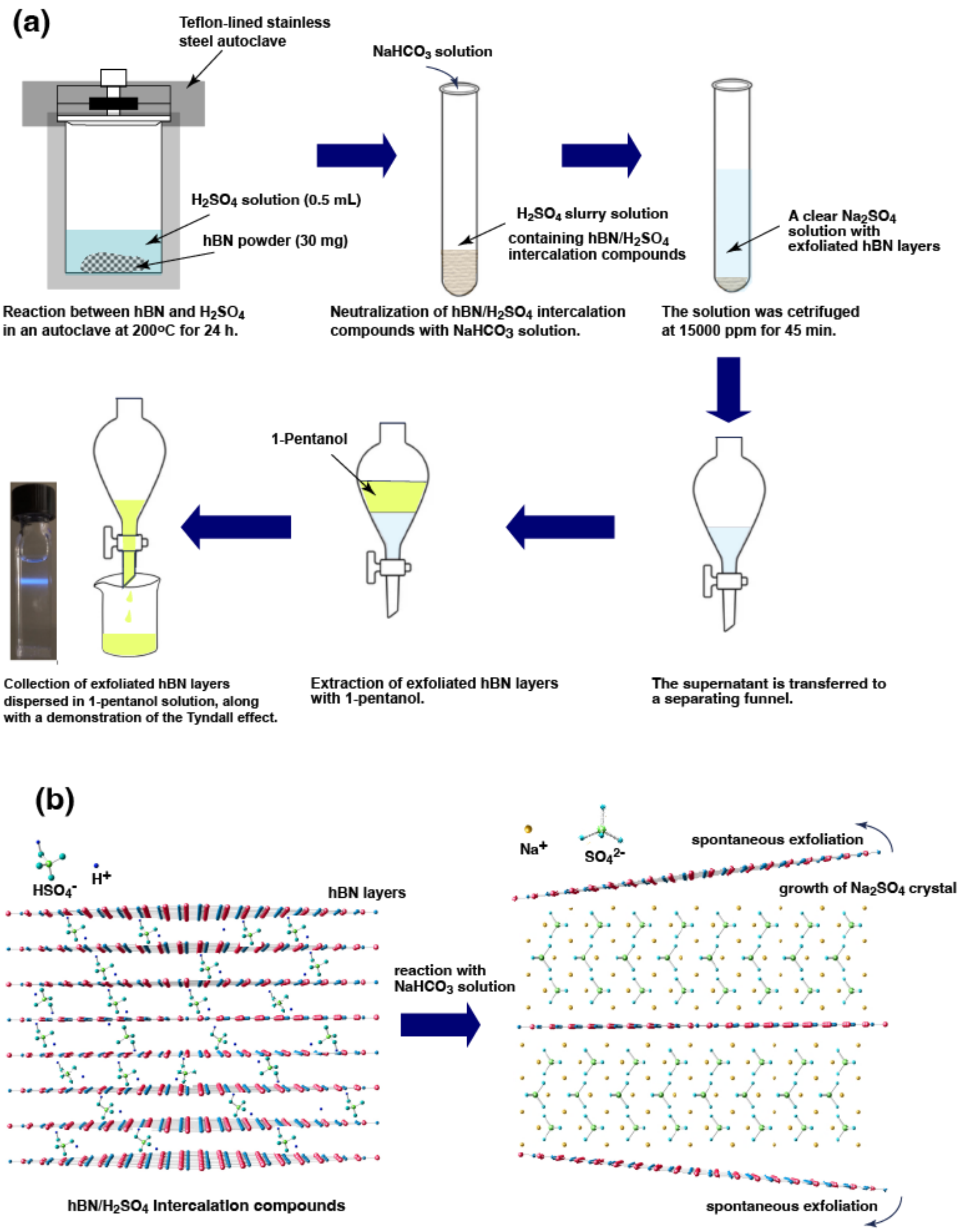

**Figure S1.** Exfoliation of hBN layers from hBN powders. (a) Formation of $hBN/H_2SO_4$ intercalation compounds, exfoliation of hBN layers via neutralization with $NaHCO_3$, and subsequent extraction of exfoliated hBN layers with 1-pentanol. (b) Microscopic scheme of the exfoliation process occurring spontaneously under the neutralization reaction.

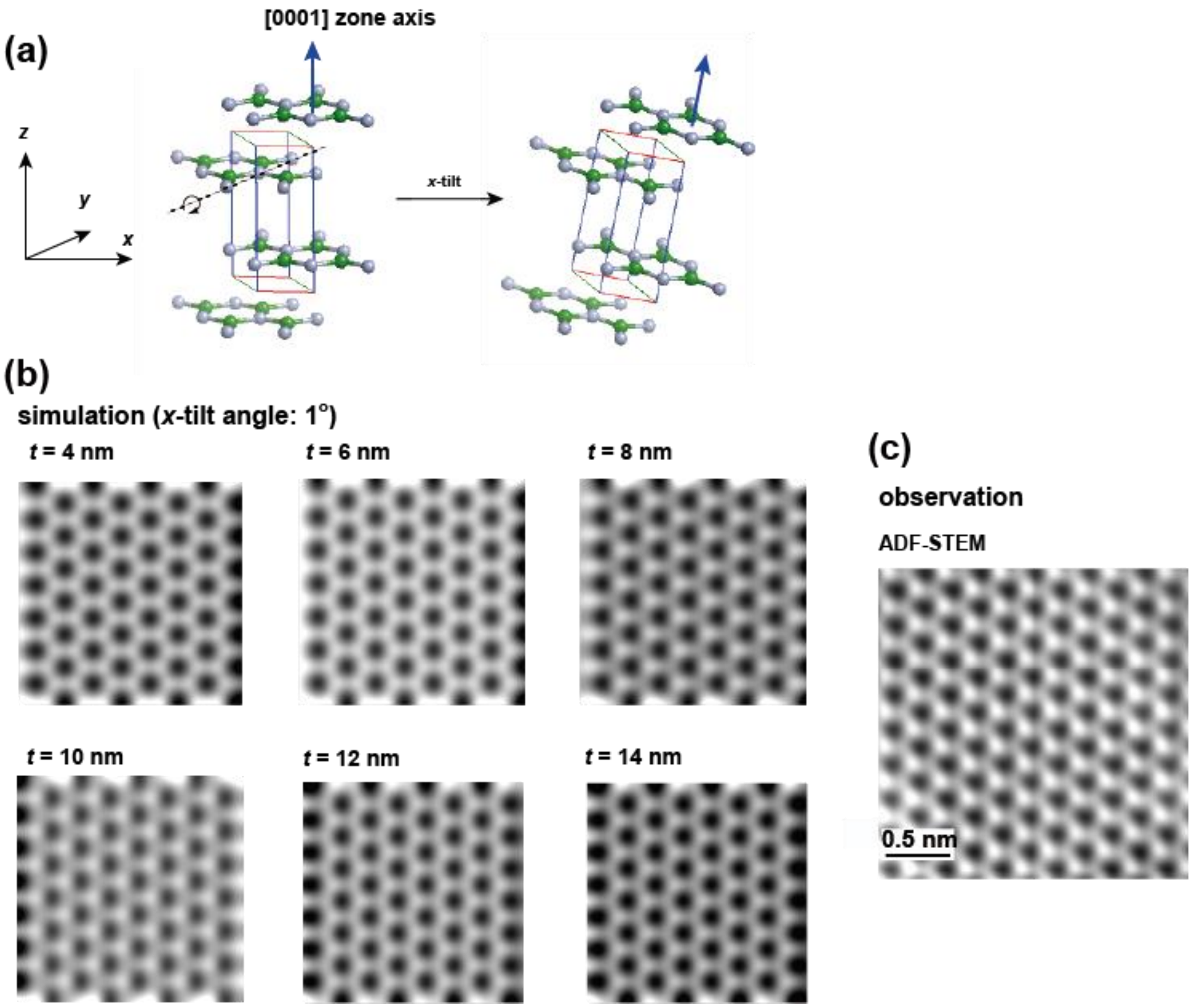

**Figure S2.** (a) Schematic illustration of an *x*-tilt performed on a hBN 0001 plane. (b) Series of simulated ADF-STEM images for the *x*-tiled (*x*-tilt angle: 1°) hBN layers with a different thickness ranging from 4 to 14 nm. (c) An example of the observed ADF-STEM image of the exfoliated specimen.

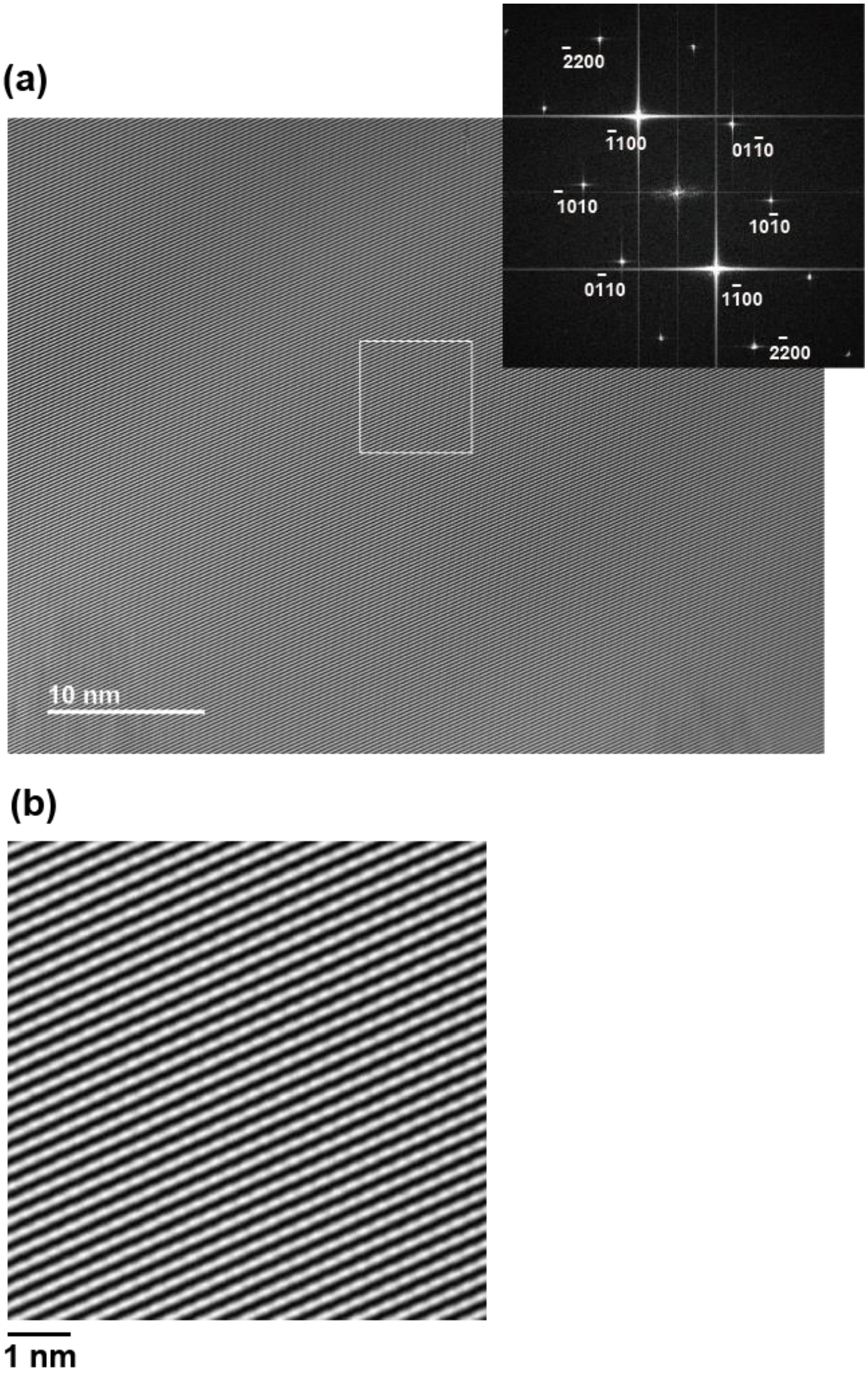

**Figure S3.** HRTEM images of non-twisted hBN layer. (a) Raw HRTEM image, along with a FFT image (inset) and (b) a magnified HRTEM image of the area marked by the white, dashed square in (a).

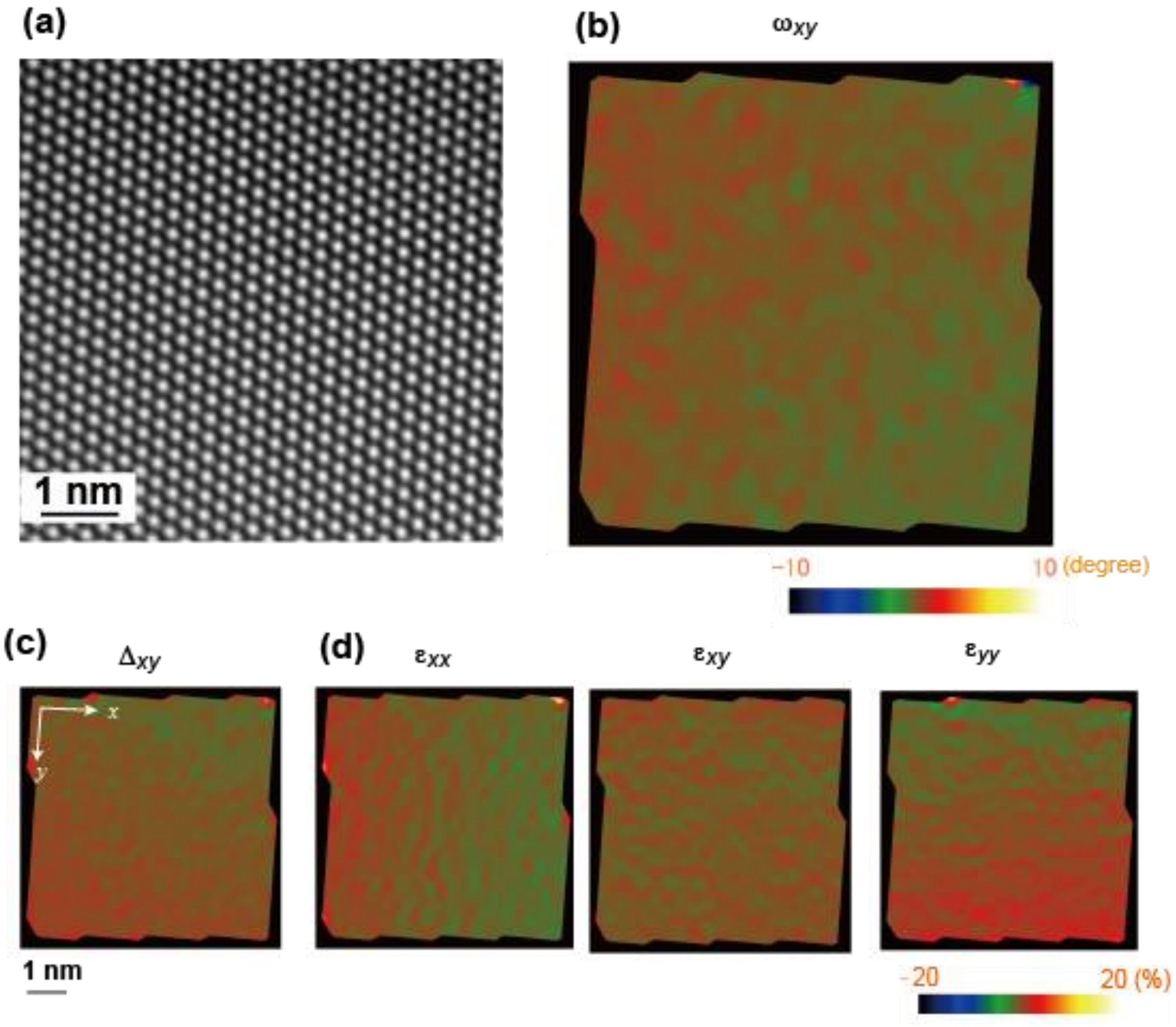

**Figure S4.** PPA analysis. (a) HRTEM image used to perform the PPA analysis. (b) Rotation (in degree and anti-clockwise positive), (c) Mean dilatation, (d) Strain tensor. The PPA analysis demonstrates that rotation is within the range of ±1 degree, whereas mean dilatation and strain tensor within the range of ±1 %.

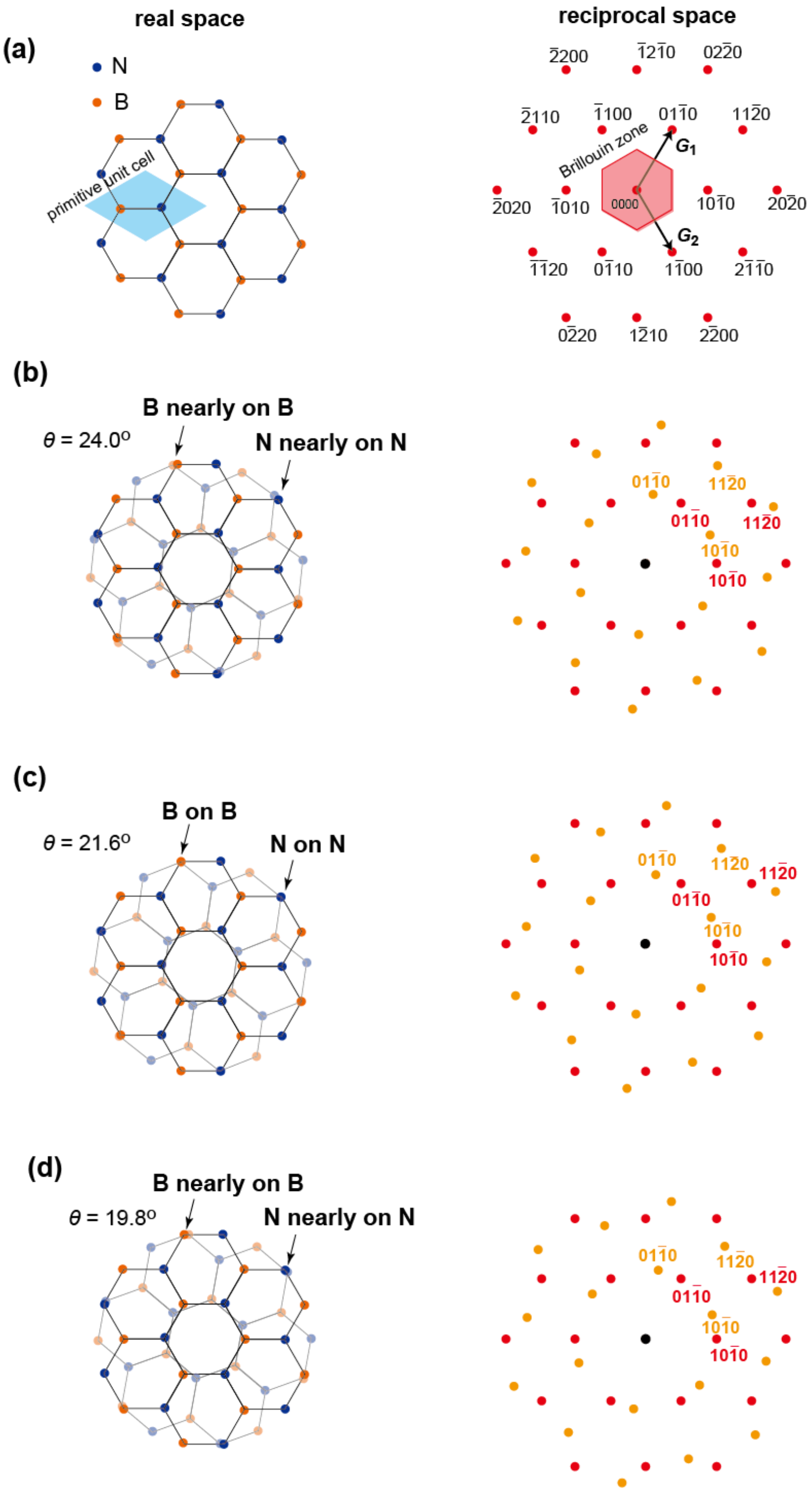

**Figure S5.** Real and reciprocal space representations of 2D hexagonal bilayers with different twist angles in the rigid lattice model. (a) hBN monolayer (left panel) and the corresponding diffraction patterns (right panel). $\boldsymbol{G}_1$ and $\boldsymbol{G}_2$ are reciprocal lattice vectors. (b–d) Atomic configurations (left panels) of twisted hBN bilayers with twist angles of (b) 24.0º, (c) 21.6º, and (d) 19.8º, and the corresponding diffraction patterns (right panels). The Miller indices, hkl, are shown to indicate the set of lattice planes responsible for each diffraction peak. Even in the incommensurate cases in (b,d), the partially eclipsed interlayer atomic overlap can be recognized.

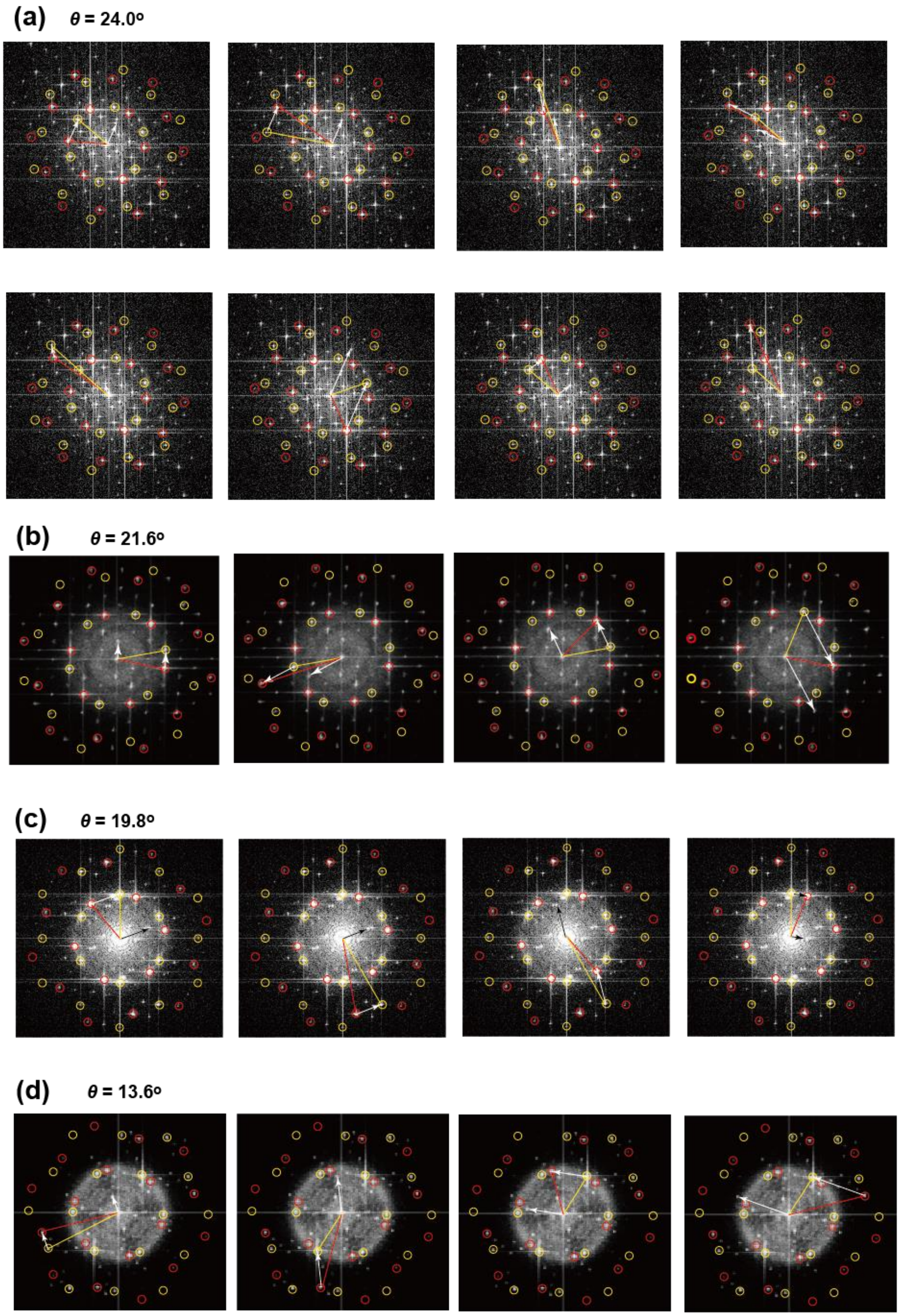

**Figure S6.** Identification of moiré peaks in FFT. (a–d) FFT images of the hBN layers with different twist angles of (a) 24.0°, (b) 21.6°, (c) 19.8°, and (d) 13.6°. Red and yellow circles show sets of $10\bar{1}0$, $11\bar{2}0$ and $20\bar{2}0$ diffraction spots belonging respectively to different layers. White or black vectors are created by connecting the two diffraction spots belonging to the different layers. These vectors provide the positions of the moiré peaks when translated to the center of the diffraction pattern.

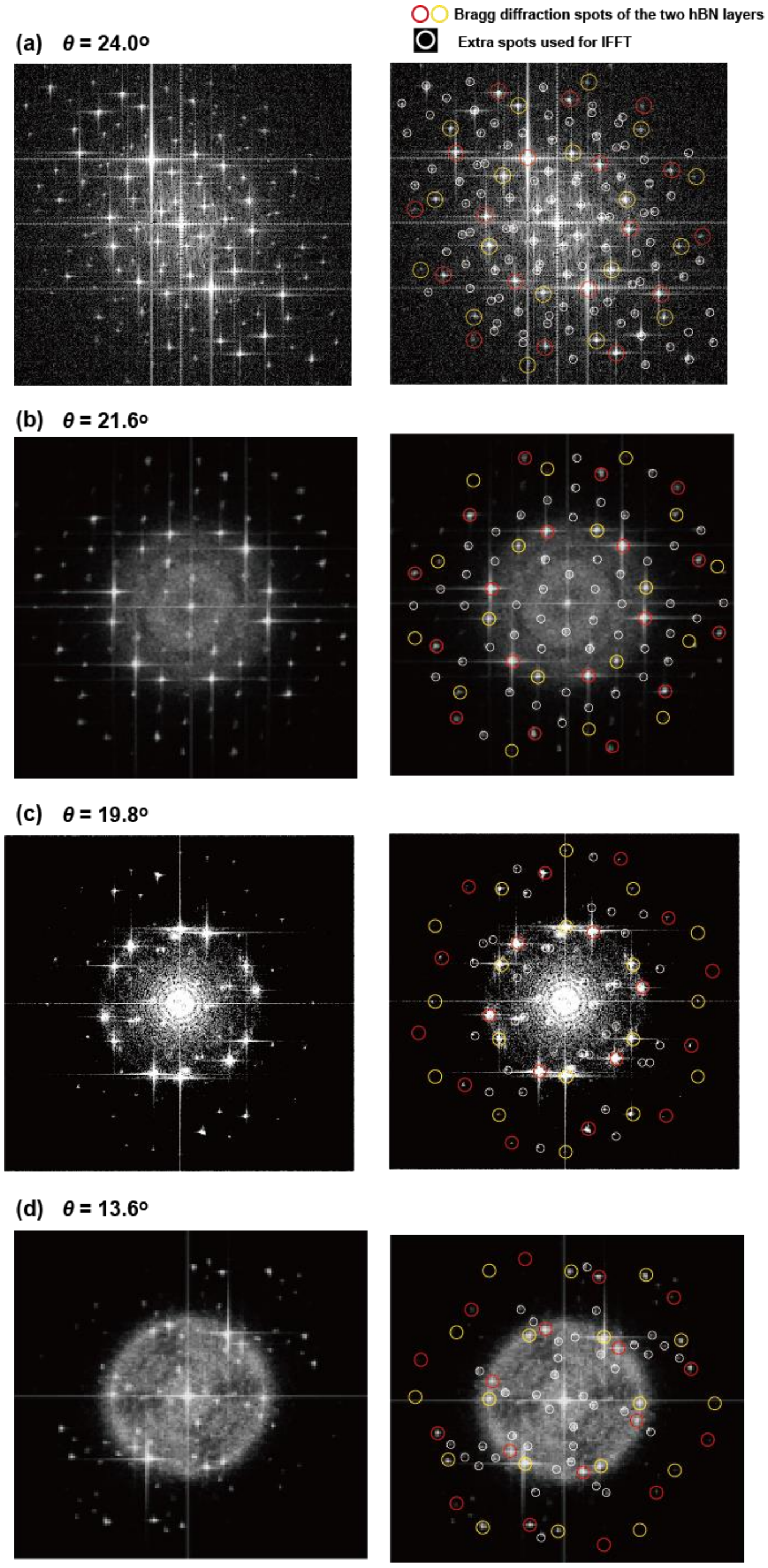

**Figure S7.** Fourier filtering process to reconstruct the moiré potential in real space by inverse FFT (IFFT). Left panels show original FFT images for the sample with a twist angle of (a) 24.0°, (b) 21.6°, (c) 19.8°, and (d) 13.6°. White circles in right panels indicate the diffraction spots used for the inclusive mask in the IFFT process. Red and yellow circles in right panels show sets of $10\bar{1}0$, $11\bar{2}0$ and $20\bar{2}0$ diffraction spots belonging respectively to different layers. These well-defined Bragg diffraction spots as well as the center spot were excluded in the IFFT process.

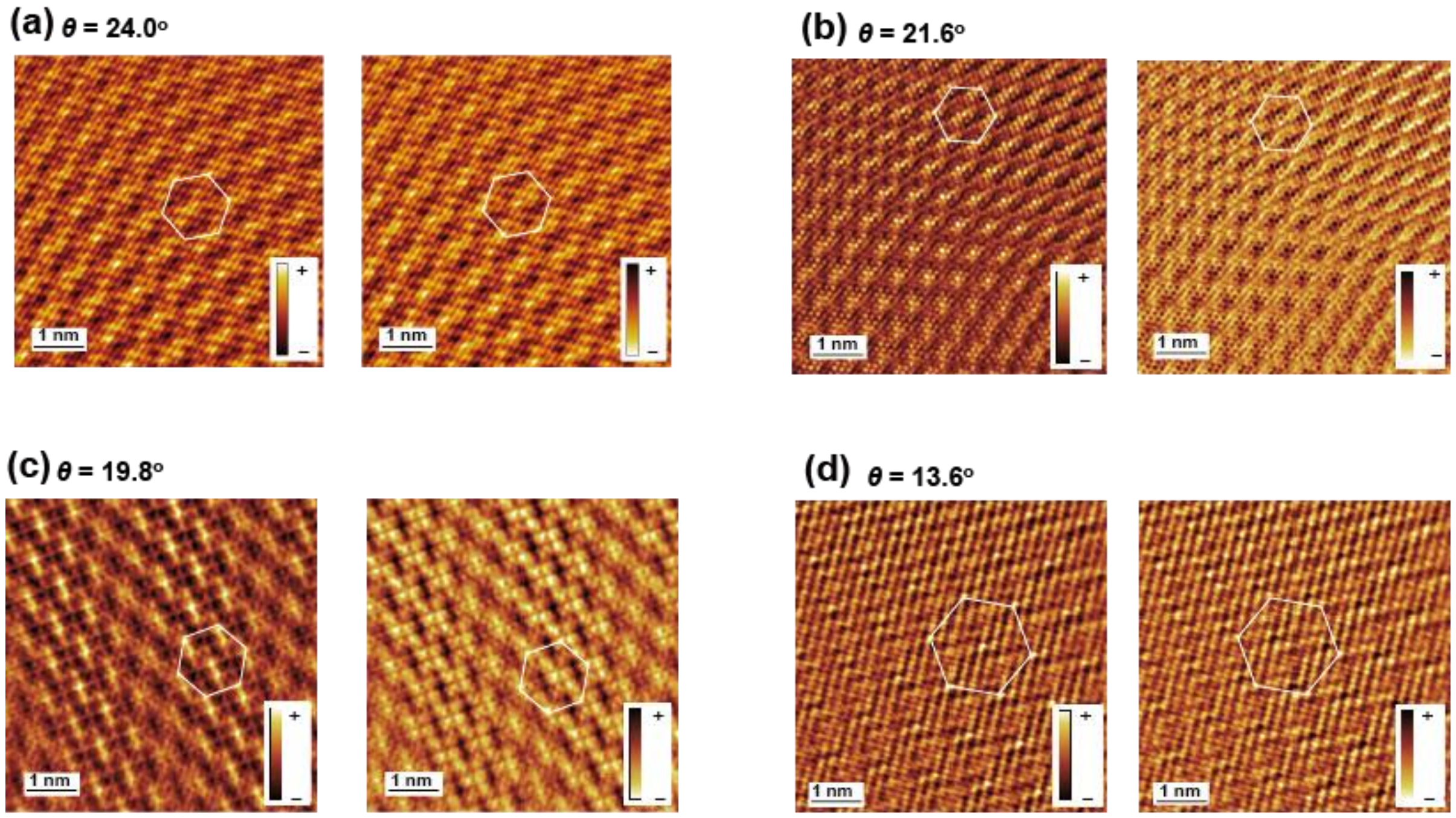

**Figure S8.** Comparison of the IFFT images between the normal and inverted contrast conditions. (a–d) IFFT images of the hBN layers with different twist angles of (a) 24.0°, (b) 21.6°, (c) 19.8°, and (d) 13.6°. Left and right panels show the normal and inverted intensities, respectively.

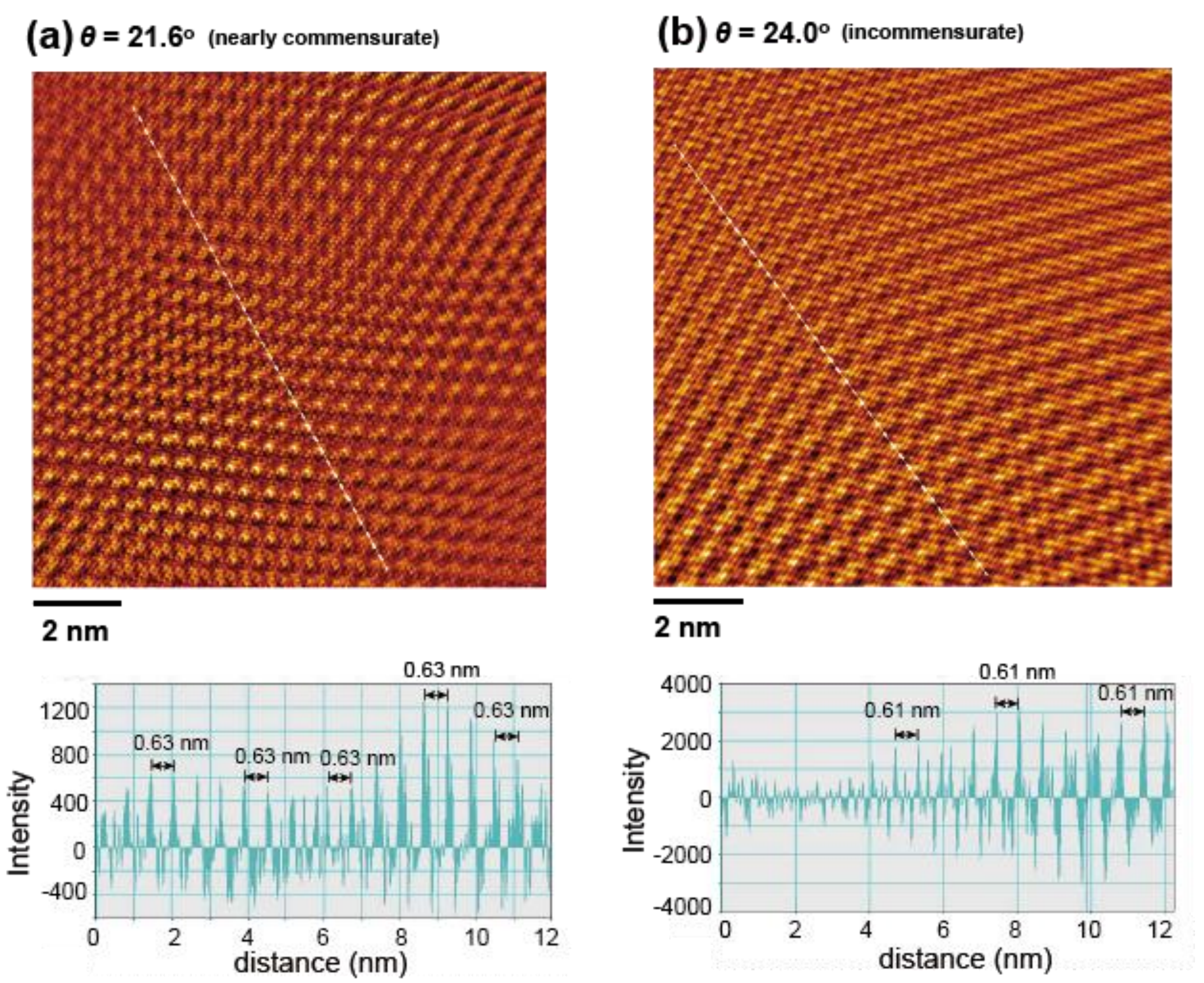

**Figure S9.** Comparison of the wide-area IFFT images between the commensurate and incommensurate layers. (a,b) (upper panels) IFFT images of the hBN layers with different twist angles of (a) 21.6° and (b) 24.0°. (lower panels) Line profiles as a function of distance along the white broken line in the corresponding upper panel.

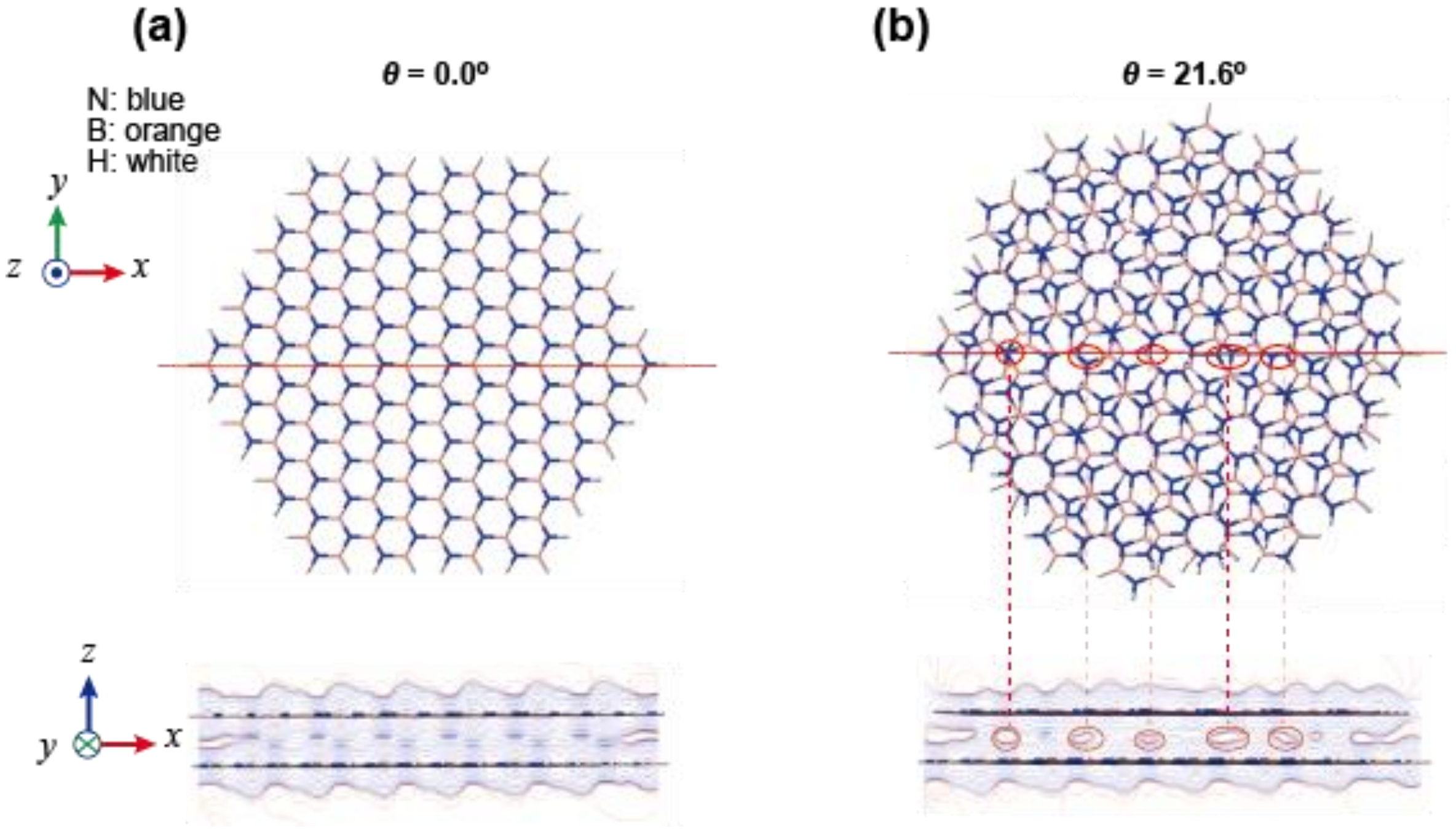

**Figure S10.** Cross-sectional ESP map perpendicular to the atomic planes. (a,b) (top panel) Optimized structure of the model clusters with twist angles of (a) $\theta = 0.0º$ and (b) 21.6º**.** (bottom panel) The corresponding cross-sectional ESP contour map along the red line shown in the top panel. The blue and pink lines are positive and negative, respectively. Contour intervals are drawn at 0.002 $e$ (a.u.)$^{-3}$. In the bottom panel in (b), negative potential regions are marked by red circles, which correspond to the N on N (or N nearly on N) regions shown in the top panel.

.

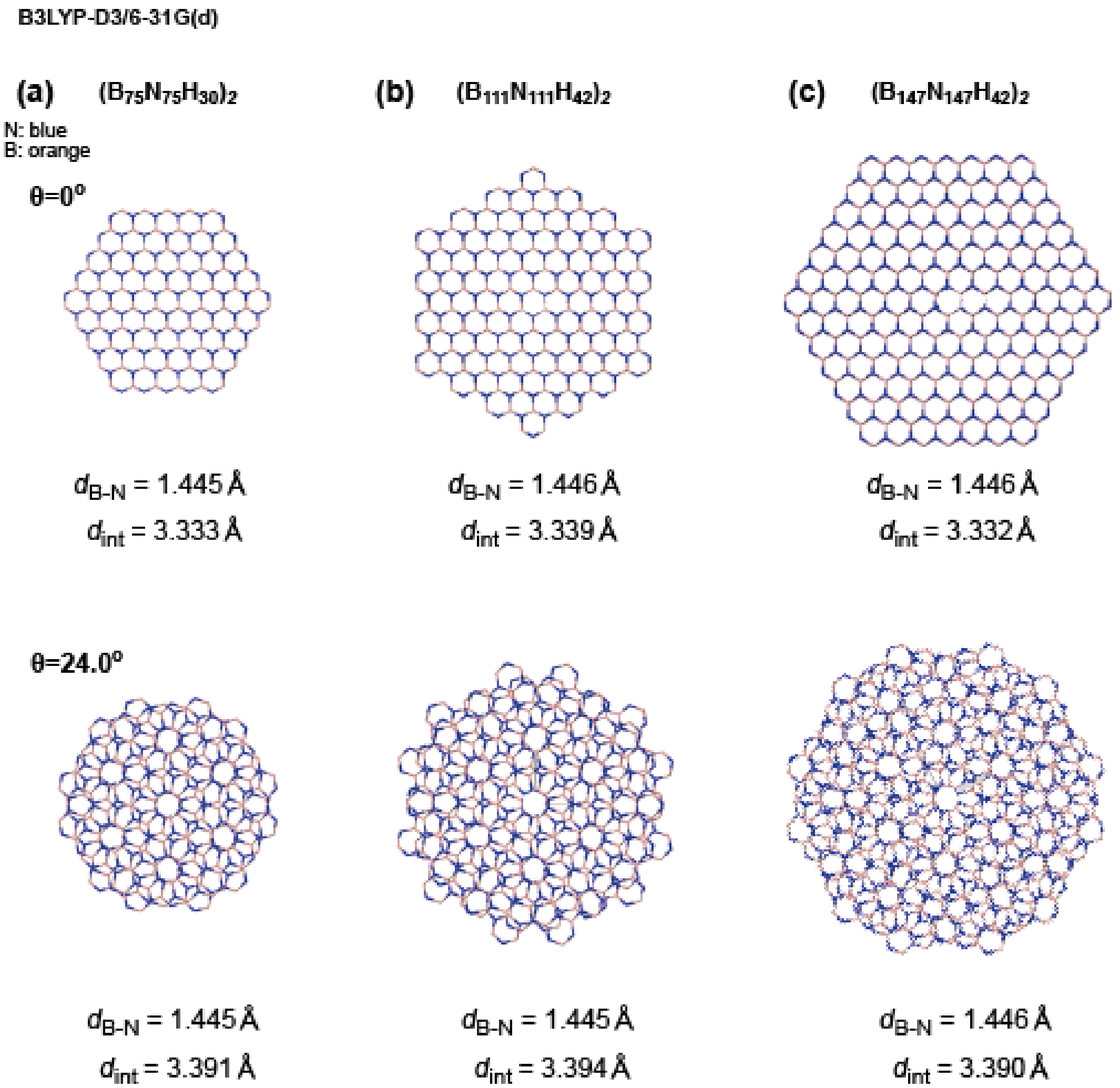

**Figure S11.** Comparison of the optimized structural parameters obtained for the three BN bilayer model clusters with different size and different twist angles ($\theta$ = 0° and 24.0°) calculated at the B3LYP-D3/6-31G(d) level. (a) $(B_{75}N_{75}H_{30})_2$ with zigzag edges. (b) $(B_{111}N_{111}H_{42})_2$ with armchair edges. (c) $(B_{147}N_{147}H_{42})_2$ with zigzag edges. The optimized parameters for the B–N bond distance $d_{B–N}$ and the interlayer distance $d_{int}$ are given. Schematic representations of the clusters illustrate the optimized geometry although the terminal H atoms are omitted for clarity.

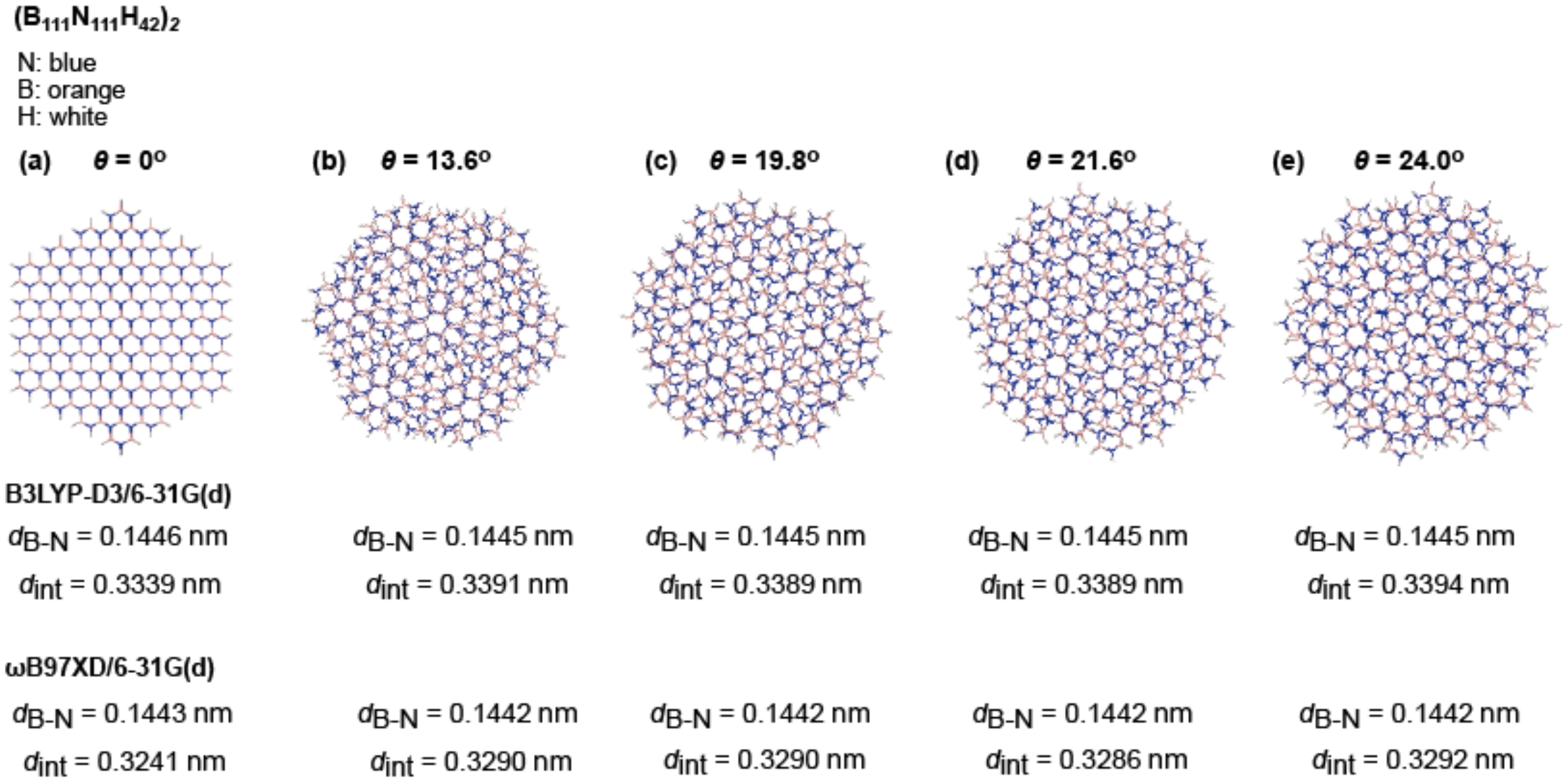

**Figure S12.** Density functional dependence of the optimized structural parameters calculated for the $(B_{111}N_{111}H_{42})_2$ clusters with different twist angles. (a) $\theta = 0°$. (b) $\theta = 13.6°$, (c) $\theta = 19.8°$, (d) $\theta = 21.6°$, and (e) $\theta = 24.0°$.

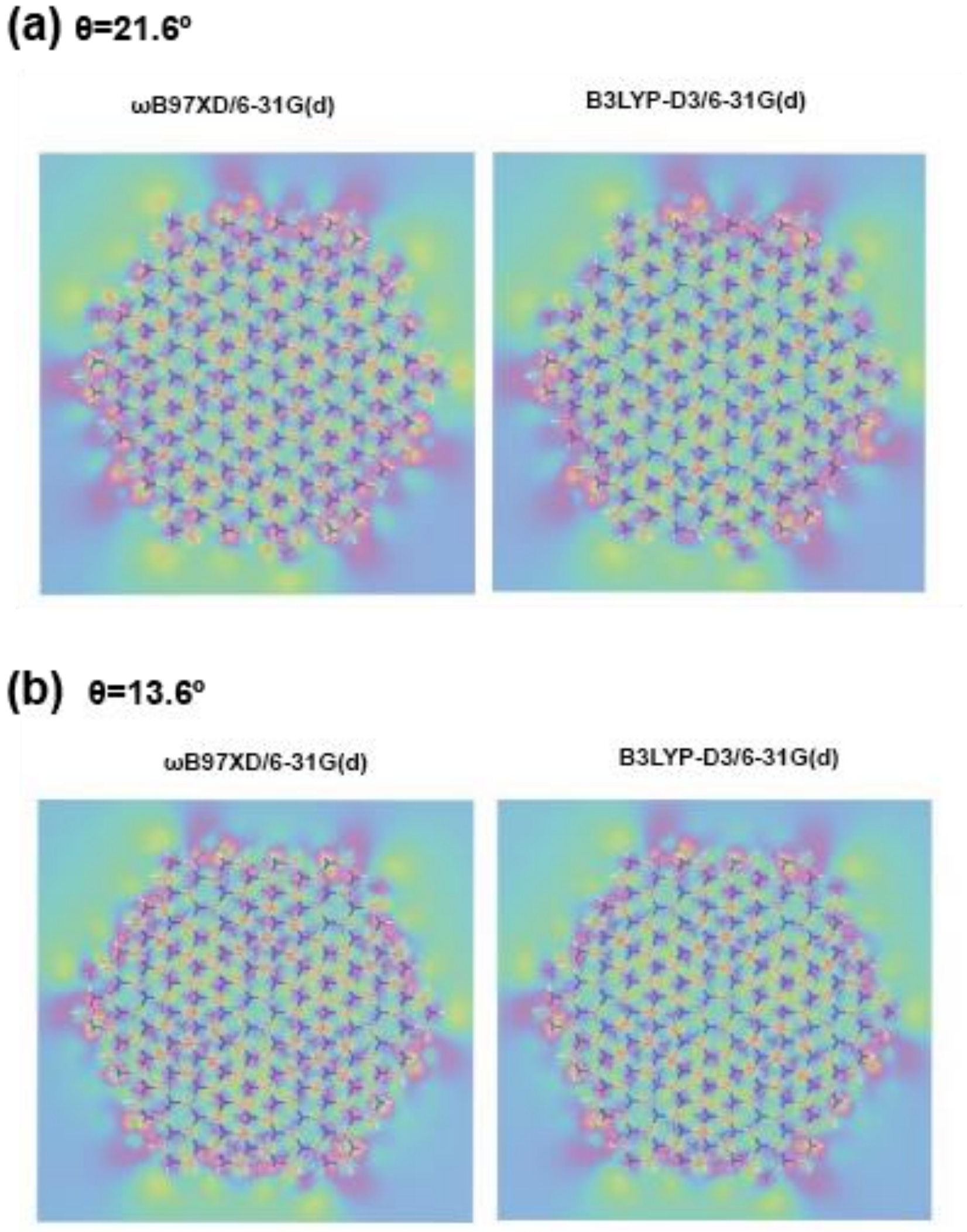

**Figure S13.** Density functional dependence of the electrostatic potential for the $(B_{111}N_{111}H_{42})_2$ clusters with different twist angles. (a) $\theta$ = 21.6°, (b) $\theta$ = 13.6°. The 2D plot of the ESP is shown for a plane bisecting the two hBN layers of the respective clusters. The unit of the scale bar is $e$ a.u.$^{-1}$.

## S6. Supporting references

(58) Zobelli, A.; Gloter, A.; Ewels, C. P. ; Seifert, G.; Colliex, C. Electron Knock-on Cross Section of Carbon and Boron Nitride Nanotubes. *Phys. Rev. B* **2007**, *75* (24), 245402.

(59) Jeon, J.W.; Kim, H.; Kim, H.; Choi, S.; Kim, B. H. Experimental Evidence for Interlayer Decoupling Distance of Twisted Bilayer Graphene. *AIP Advances* **2018**, *8* (7), 075228.

(60) Liao, M.; Wei, Z.; Du, L.; Wang, Q.; Tang, J.; Yu, H.; Wu, F.; Zhao, J.; Xu, X.; Han, B.; Liu, K.; Gao, P.; Polcar, T.; Sun, Z.; Shi, D.; Yang, R.; Zhang, G. Precise Control of the Interlayer Twist Angle in Large Scale $MoS_2$ Homostructures. *Nat. Commun*. **2020**, *11*, 2153.

(61) Kamiya, M.; Tsuneda, T.; Hirao, K. A Density Functional Study of van der Waals Interactions. *J. Chem. Phys.* **2002**, *117* (13), 6010–6015.

(62) Becke, A. D. Density-Functional Thermochemistry. III. The Role of Exact Exchange. *J. Chem. Phys*. **1993**, *98* (7), 5648–5652.

(63) Grimme, S.; Antony, J.; Ehrlich, S.; Krieg, H. A Consistent and Accurate ab initio Parameterization of Density Functional Dispersion Correction (DFT-D) for the 94 elements H-Pu. *J. Chem. Phys.* **2010**, *132* (15), 154104.

(64) Chai, J.-D.; Head-Gordon, M. Long-Range Corrected Hybrid Density Functionals with Damped Atom–Atom Dispersion Corrections. *Phys. Chem. Chem. Phys.* **2008**, *10* (44), 6615-6620.

(65) Pease, R. S. Crystal structure of boron nitride. *Nature* **1950**, *165*, 722–723.

(66) Warner, J. H.; Rümmel, M. H.; Bachmatiuk, A.; Büchner, B. Atomic Resolution Imaging and Topography of Boron Nitride Sheets Produced by Chemical Exfoliation. *Nano Lett.* **2010**, *4* (3), 1299-1305.

(67) Jensen, F. *Introduction to Computational Chemistry*, 3rd ed.; Wiley: Chichester, 2017; p 321.

(68) Humphrey, W.; Dalke, A;. Schulten, K. VMD: Visual Molecular Dynamics. *J. Mol. Graph*. **1996**, *14* (1), 33–38.

(69) Frisch, A.; Nielsen, A. B.; Holder, A. J. *Gaussview Users Manual*; Gaussian Inc.: Pittsburgh, PA, 2000.